\documentclass{ptephy_v1}

\newtheorem{definition}{Definition}

\newtheorem{conjecture}{Conjecture}

\begin{document}

\title{Formation of dynamically transversely trapping surfaces and the stretched hoop conjecture}



\author{Hirotaka Yoshino${}^1$}
\author{Keisuke Izumi${}^{2,3}$}
\author{Tetsuya Shiromizu${}^{3,2}$}
\author{Yoshimune Tomikawa${}^4$}
\affil{${}^1$Advanced Mathematical Institute, Osaka City University, Osaka 558-8585, Japan}
\affil{${}^2$Kobayashi-Maskawa Institute, Nagoya University, Nagoya 464-8602, Japan}
\affil{${}^3$Department of Mathematics, Nagoya University, Nagoya 464-8602, Japan}
\affil{${}^4$Faculty of Economics, Matsuyama University, Matsuyama 790-8578, Japan}


\begin{abstract}
  A dynamically transversely trapping surface (DTTS)
  is a new concept of an extension of a photon sphere
  that appropriately represents a strong gravity region and
  has close analogy with a trapped surface.
  We study formation of
  a marginally DTTS in 
  time-symmetric, conformally flat initial
  data with two black holes, with a spindle-shaped source, and
  with a ring-shaped source, and 
  clarify that $\mathcal{C}\lesssim 6\pi GM$ describes the
  condition for the DTTS formation well, where $\mathcal{C}$ is the circumference
  and $M$ is the mass of the system.
  This indicates that an 
  understanding analogous to the hoop conjecture
  for the horizon formation is possible.
  Exploring the ring system further, we find configurations where 
  a marginally DTTS with the torus
  topology forms inside a marginally
  DTTS with the spherical topology,
  without being hidden by an apparent horizon.
  There also exist configurations where
  a marginally trapped surface
  with the torus topology 
  forms inside a marginally trapped surface with the spherical topology,
  showing a further similarity between
  DTTSs and trapped surfaces.
\end{abstract}

\subjectindex{E00, E31, A13}

\maketitle

%
\section{Introduction}
\label{section1}

The recent observation 
of a black hole shadow \cite{EHTCollaboration:2019}
produced by a central massive object
in the galaxy M87 significantly increased the importance of the concept of
a photon sphere \cite{Virbhadra:1999}.
A photon sphere is a spherically symmetric surface on which
circular orbits of photons exist, and 
in a Schwarzschild black hole, it is located at $r=3GM$
where $r$ is the areal radius, $G$ is the Newtonian constant
of gravitation, and $M$ is the Arnowitt--Deser--Misner mass that represents
the total energy of the black hole.   
The edge of a black hole shadow is primarily
determined by the photon sphere, or its extension, the fundamental
photon orbits \cite{Cunha:2017}.
Similarly to event/apparent horizons as extended concepts of $r=2GM$
of a Schwarzschild spacetime, appropriately extended concepts
of a photon sphere $r=3GM$ would also significantly advance 
our understanding of spacetimes with strong gravity regions. 
Several extended concepts of a photon sphere have been proposed
so far: a photon surface \cite{Claudel:2000},
a loosely trapped surface \cite{Shiromizu:2017},
a static/stationary transversely trapping surface (TTS) \cite{Yoshino:2017},
a wandering set \cite{Siino:2019},  
a dynamically transversely trapping surface (DTTS) \cite{Yoshino:2019},
and a quasi-local photon surface \cite{Cao:2019}.
In this paper we focus attention on a DTTS
proposed in our previous paper. 
A (marginally) DTTS is an analogous concept  
to a (marginally) trapped surface, appropriately represents
a strong gravity region outside a horizon, and  
is easily calculated.

In our previous paper \cite{Yoshino:2019} we
pointed out similarities between 
a (marginally) DTTS and a (marginally) trapped surface.
Both surfaces are determined on a spacelike hypersurface
and have similar gauge dependence properties.
In time-symmetric initial data,
the area $A_0$ of a convex DTTS satisfies the Penrose-like inequality
$A_0\le 4\pi (3GM)^2$, similarly to the fact that the area $A_{\rm AH}$ of an
apparent horizon satisfies the Riemannian Penrose inequality
$A_{\rm AH}\le 4\pi (2GM)^2$ \cite{Wald:1977,Huisken:2001,Bray:2001},
which is a special case of the Penrose conjecture \cite{Penrose:1973}. 
We explore the similarity between the two surfaces
further in this paper, paying attention to the
condition for the formation of the two kinds of surfaces.
As the condition for the horizon formation, the
hoop conjecture has been proposed by Thorne \cite{Thorne:1972}:
\begin{conjecture}
Black holes with horizons form when and only when a mass $M$
gets compacted into a region whose circumference in every direction
is bounded by $\mathcal{C}\lesssim 4\pi GM$.
\end{conjecture}
\noindent
Here, $4\pi GM$ is the circumference of the horizon
of a Schwarzschild black hole with a mass $M$, i.e. $2\pi (2GM)$.
Although the hoop conjecture is loosely formulated,
it is tested in many examples and the results basically
support this conjecture \cite{Nakamura:1988,Wojtkiewicz:1990,Shapiro:1991,Flanagan:1991,Barrabes:1991,Tod:1992,Brrabes:1992,Flanagan:1992,Abrahams:1992,Chiba:1994,Bernstein:1994,Chiba:1998,Ida:1998,Yoshino:2001,Yoshino:2007,Choptuik:2009,Rezzolla:2012,East:2019}. 
One implication of this conjecture is that
an apparent horizon which is arbitrarily long in one direction
does not form.
Similarly, the condition $\mathcal{C}\lesssim 6\pi GM$
may be expected as the condition for the formation of a marginally DTTS,
where $6\pi GM$ is the circumference of a photon sphere
of a Schwarzschild black hole with mass $M$, i.e. $2\pi (3GM)$.

The hoop conjecture is also related to topological properties
of apparent horizons.
In higher-dimensional spacetimes, it is known that horizons of
stationary black holes 
can have nonspherical topologies like black strings
or black rings \cite{Emparan:2001,Pomeransky:2006}.
This is understood from the fact that
the hoop conjecture does not hold in higher-dimensional
spacetimes. In Ref.~\cite{Ida:2002},
initial data with a
spindle-shaped source in a five-dimensional spacetime
are studied, and an apparent horizon which is arbitrarily long 
in one direction is shown to form (see also 
Refs.~\cite{Nakao:2001,Barrabes:2004,Yoo:2005,Yamada:2009,Cvetic:2011,Kurata:2012,Mujtaba:2012}
for related studies).
From the formation of an arbitrarily long apparent horizon,
the formation of an apparent horizon with the topology $S^1\times S^2$
is also expected by slightly bending and connecting the edges of
a long apparent horizon. In fact, initial data with a ring-shaped source
are also studied in Ref.~\cite{Ida:2002},
and an apparent horizon with the topology $S^1\times S^2$ forms
for a sufficiently large radius of the ring.
Conversely, in four-dimensional spacetimes,
since an apparent horizon cannot be long in one
direction due to the hoop conjecture, an apparent horizon
with the torus topology is not expected to form \cite{Yoshino:2007}.
More precisely, even if a marginally trapped surface
with the torus topology (a {\it marginally trapped torus}, hereafter)
forms, 
it would be surrounded by a marginally trapped surface
with the spherical topology (hereafter, {\it a marginally trapped sphere}).
In fact, all existing examples of marginally trapped tori
in asymptotically flat initial data are
hidden by marginally trapped spheres \cite{Husa:1996,Karkowski:2017}.

Motivated by the above discussions, we study two issues in this paper.
First, we examine whether the understanding that is analogous
to the hoop conjecture is possible or not for the formation
of a marginally DTTS with the spherical topology
(hereafter, a {\it marginally DTT sphere}).
For this purpose, we study time-symmetric conformally
flat initial data with two black holes, with a spindle-shaped source,
and with a ring-shaped source. In particular,
we will show that a marginally DTT sphere that is arbitrarily
long in one direction does not form in the spindle initial data.
We also show that in all systems, the condition $\mathcal{C}\lesssim 6\pi GM$
reasonably gives the necessary and sufficient conditions
for the formation of (outermost) marginally DTT spheres.
We will call this condition the {\it stretched hoop conjecture}.

Next, we study the formation of a marginally DTTS with the torus topology
(hereafter, a {\it marginally DTT torus}) by studying the ring system
in more detail. The stretched hoop conjecture indicates that
if a marginally DTT torus forms, it would be located inside a
marginally DTT sphere. In the ring system, it will be demonstrated 
that there is a parameter region where a marginally DTT torus
forms, and a marginally DTT torus is always located
inside a marginally DTT sphere at least in the ring system.
Note that although we proved in our previous paper \cite{Yoshino:2019}
that a convex DTTS must have the spherical topology,
the formation of marginally DTT tori here does not
contradict the theorem because they are not convex. 
Furthermore, we provide a further similarity between marginally DTTSs
and marginally trapped surfaces by showing that
there is a parameter region where a marginally trapped torus
forms inside a marginally trapped sphere.

This paper is organized as follows.
In the next section we briefly review the
definition of a (marginally) DTTS and present the equations 
to solve for a marginally DTTS in time-symmetric initial data.
In Sect.~\ref{section3}, we explain the setup of the three systems
studied in this paper, i.e. two-black-hole initial data,
spindle initial data, and ring initial data. The numerical 
method of solving for a marginally DTTS in these systems
is also explained. In Sect.~\ref{section4} we examine whether
the understanding that is analogous to the hoop conjecture
for horizon formation is possible or not
for marginally DTT spheres.
In Sect.~\ref{section5} the formation of marginally DTT tori
is examined in the ring system.
Also, the formation of marginally trapped spheres/tori
is examined to show the similarity between marginally DTTSs
and marginally trapped surfaces.
Section~\ref{section6} is devoted to a summary and discussions.
In Appendix~\ref{Appendix-A}, the numerical method
of solving for marginally DTT tori in the ring system
is explained. In Appendix~\ref{Appendix-B}, the formation
of marginally DTT tori is studied approximately
in the situation where the ring radius $R$ is much
smaller than $GM$. 
Throughout the paper, 
we work in the framework of the theory of general relativity
for four-dimensional spacetimes.  
We use units in which the speed of light is unity, $c=1$,
while the Newtonian constant of gravitation $G$ is explicitly shown.

%
%
\section{Dynamically transversely trapping surfaces}
\label{section2}

In this section we present a brief review of the
definition of a (marginally) DTTS that was proposed
in our previous paper \cite{Yoshino:2019}. The definition
is given in Sect.~\ref{section2-1}, and useful formulae
to solve for a marginally DTTS 
in time-symmetric initial data are presented in Sect.~\ref{section2-2}.
We refer readers to our previous paper \cite{Yoshino:2019}
for more detailed explanations and derivations.

\subsection{Definition}
\label{section2-1}

%
\begin{figure}[tb]
\centering
\includegraphics[width=0.7\textwidth,bb=0 0 401 237]{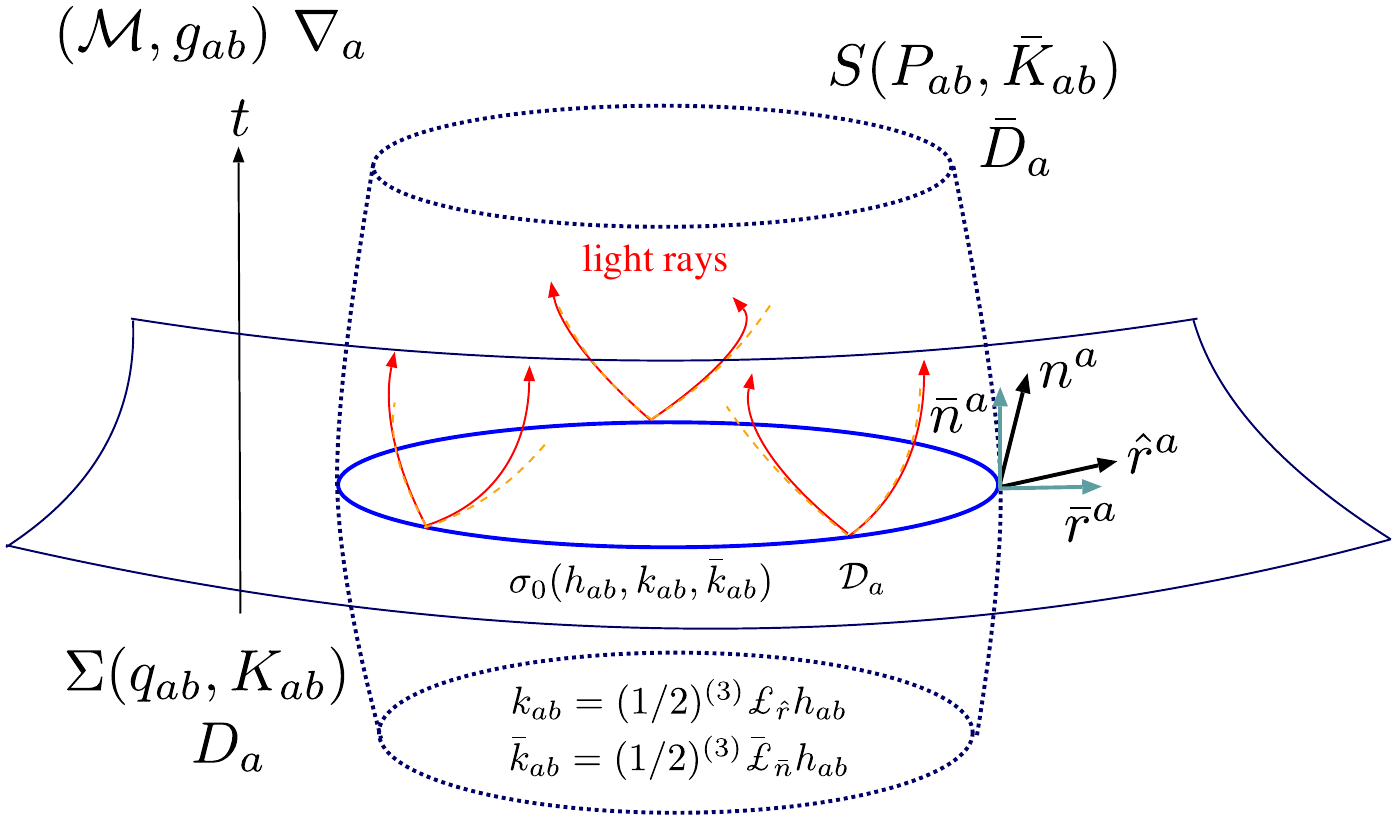}
\caption{
  Configuration to be considered.
  A two-dimensional closed surface $\sigma_0$ exists in a
  spacelike hypersurface $\Sigma$ of a spacetime $\mathcal{M}$.
  A timelike hypersurface $S$ intersects with $\Sigma$
  precisely at $\sigma_0$. 
  Notations are also indicated.
  See text and Ref.~\cite{Yoshino:2019} for details.
}
\label{schematic-DTTS}
\end{figure}
%

Figure~\ref{schematic-DTTS} depicts the configuration
to be considered. 
Let us consider a spacelike hypersurface $\Sigma$ 
with a future-directed
unit normal $n^a$ in a spacetime $\mathcal{M}$
with a metric $g_{ab}$.
The induced metric and the extrinsic curvature of $\Sigma$
are $q_{ab}=g_{ab}+n_an_b$ and $K_{ab} = (1/2)\pounds_n q_{ab}$,
respectively, 
where $\pounds$ is a Lie derivative with respect to $\mathcal{M}$.
A DTTS is a two-dimensional closed orientable 
surface $\sigma_0$ in a spacelike hypersurface $\Sigma$.
The two-dimensional surface $\sigma_0$ can be regarded
as a surface in $\Sigma$, and the outward unit normal
to $\sigma_0$ in this sense is denoted by $\hat{r}^a$.
The induced metric and the extrinsic curvature of $\sigma_0$
(as a surface in $\Sigma$) are 
$h_{ab}=q_{ab}-\hat{r}_a\hat{r}_b$
and $k_{ab} = (1/2){}^{(3)}\pounds_{\hat{r}}h_{ab}$, respectively,
where ${}^{(3)}\pounds$ is a Lie derivative with respect to $\Sigma$.
We introduce a timelike hypersurface $S$ in $\mathcal{M}$, which
intersects with $\Sigma$ precisely at $\sigma_0$.
Denoting the outward unit normal to $S$ as $\bar{r}^a$,
the induced metric and the extrinsic curvature of $S$
are $\bar{p}_{ab}=g_{ab}-\bar{r}_a\bar{r}_b$
and $\bar{K}_{ab} = (1/2)\pounds_{\bar{r}}\bar{p}_{ab}$, respectively.
The two-dimensional surface $\sigma_0$ can be regarded
as a surface in $S$, and the future-directed unit normal
to $\sigma_0$ that is tangent to $S$ is denoted by $\bar{n}^a$.
Note that the two hypersurfaces 
$S$ and $\Sigma$ are not necessarily orthogonal to each other,
and hence $\bar{r}^a$ and $\bar{n}^a$ do not
coincide with $\hat{r}^a$ and $n^a$ in general, respectively.
The extrinsic curvature of $\sigma_0$ as a surface in $S$
is defined by $\bar{k}_{ab} = (1/2){}^{(3)}\bar{\pounds}_{\bar{n}}h_{ab}$,
where ${}^{(3)}\bar{\pounds}_{\bar{n}}$ is a Lie derivative
associated with $S$.

With these notations, a DTTS is defined as follows: 
%
\begin{definition}
  Suppose $\Sigma$ to be a smooth spacelike hypersurface of
  a spacetime $\mathcal{M}$.
  A smooth closed orientable two-dimensional
  surface $\sigma_0$ in $\Sigma$ is a dynamically transversely trapping
  surface (DTTS) if and only if there exists a timelike hypersurface $S$
  in $\mathcal{M}$ that intersects $\Sigma$ precisely at $\sigma_0$ 
  and satisfies the following three conditions at arbitrary points on
  $\sigma_0$:
\begin{eqnarray}
  \bar{k} = 0, &\quad&
  \textrm{(the momentarily non-expanding condition),}
  \label{momentarily-non-expanding-condition}
  \\
  \mathrm{max}\left(\bar{K}_{ab}k^ak^b\right) = 0, &\quad&
  \textrm{(the marginally transversely trapping condition),}
  \label{marginally-transversely-trapping-condition}
  \\
  {}^{(3)}\bar{\pounds}_{\bar{n}} \bar{k}\le 0, &\quad&
  \textrm{(the accelerated contraction condition),} 
  \label{accelerated-contraction-condition}
\end{eqnarray}
where $k^a$ are arbitrary future-directed null vectors tangent to $S$ and
the quantity $\pounds_{\bar{n}}\bar{k}$ is evaluated
with a time coordinate in $S$ whose lapse function is constant
on $\sigma_0$.
\label{definition-1}
\end{definition}
%

A physical interpretation for the above definition is as follows.
The first two conditions, Eqs.~\eqref{momentarily-non-expanding-condition} and 
\eqref{marginally-transversely-trapping-condition}, 
specify the structure of a timelike hypersurface $S$
up to the second order in time from the behavior of photons.
The momentarily non-expanding condition
of Eq.~\eqref{momentarily-non-expanding-condition}
means that the hypersurface $S$ is chosen so that the surface
$\sigma_0$ becomes an extremal surface in $S$. 
Then, we consider photons emitted in arbitrary tangent directions
to $S$ from arbitrary points on $\sigma_0$. 
The marginally transversely trapping condition
of Eq.~\eqref{marginally-transversely-trapping-condition}
indicates that all photons emitted tangentially to $S$
from each point
must propagate on or in an 
inward direction of $S$, and also,
at least one photon must propagate on $S$.
In other words, if we consider a collection of photons
emitted from all points on $\sigma_0$ tangentially to $S$,
they distribute in a region with small thickness,
and $S$ is adopted as the outer boundary of that region.
The accelerated contraction condition of
Eq.~\eqref{accelerated-contraction-condition}
is imposed to judge whether $\sigma_0$ is in a strong gravity region.
If $\sigma_0$ is a DTTS, the hypersurface $S$ determined by the above
procedure becomes a maximal surface for the time slices given
by the constant lapse function $\alpha$ on $\sigma_0$.

In Definition 1, we allow both non-self-intersecting 
and self-intersecting surfaces as DTTSs.
In the case of marginally outer trapped surfaces (MOTS),
self-intersection is allowed and 
explicit examples of self-intersecting MOTSs have been constructed
in a numerical simulation of a two-black-hole collision \cite{Pook-Kolb:2019}.
In a similar manner, a self-intersecting DTTS may form. 
But in the explicit construction of marginally DTTSs in
Sects.~\ref{section4} and \ref{section5} in this paper,
we focus on non-self-intersecting DTTSs for simplicity.

As an example, let us consider an $r=\mathrm{const.}$ sphere $\sigma_0$
on a $t=\mathrm{const.}$ hypersurface in a Schwarzschild spacetime.
The momentarily non-expanding condition 
of Eq.~\eqref{momentarily-non-expanding-condition}
means that we consider photons emitted with the initial condition
$dr/dt=0$ from $\sigma_0$.
Due to the spherical symmetry,
the timelike hypersurface
$S$ satisfying the marginally transversely trapping condition
of Eq.~\eqref{marginally-transversely-trapping-condition}
is the one composed of worldlines of all photons,
which corresponds to the
photon surface \cite{Claudel:2000}. Calculating
the quantity ${}^{(3)}\bar{\pounds}_{\bar{n}} \bar{k}$, we find
that an $r=\mathrm{const.}$ sphere with $r\le 3GM$ satisfies
the accelerated contraction condition, while a surface with
$r>3GM$ does not satisfy it.

We define a dynamically transversely trapping region
and a marginally DTTS with the following definition:
%
\begin{definition}
  Consider a collection of 
  all DTTSs such that 
  any two of these DTTSs can be transformed
  to one another by 
  continuous deformation without violating the three conditions
  of Eqs.~\eqref{momentarily-non-expanding-condition}--\eqref{accelerated-contraction-condition}.
  The region in which these DTTSs exist is said to be a 
  dynamically transversely trapping region (or, more generally, 
  one of the dynamically transversely trapping regions). If the
  outer boundary of a dynamically transversely trapping region
  satisfies
\begin{equation}
  \pounds_{\bar{n}} \bar{k}= 0,
  \label{Equation-for-marginally-DTTS}
\end{equation}
it is said to be a marginally DTTS.
\label{definition-2}
\end{definition}
%

\subsection{Time-symmetric case}
\label{section2-2}

Here, we briefly describe how to find
a marginally DTTS in time-symmetric initial data.
In this case, 
a timelike hypersurface $S$ that
intersects $\Sigma$ orthogonally 
satisfies $\bar{k}_{ab}=0$ at $\sigma_0$, 
and thus satisfies the momentarily non-expanding
condition, Eq.~\eqref{momentarily-non-expanding-condition}. Therefore,
the relations $\bar{r}^a=\hat{r}^a$ and $\bar{n}^a=n^a$ hold,
and the analysis is simplified.
We span the coordinates in the neighborhood of $\sigma_0$
so that the metric takes the form
\begin{equation}
  ds^2 = -\alpha^2dt^2 + \varphi^2 dr^2 + h_{ij}dx^idx^j + 2\gamma_{ri}drdx^i,
  \label{metric-neighborhood-of-sigma0}
\end{equation}
where the spacelike hypersurface $\Sigma$ is given by $t=0$,
the timelike hypersurface $S$ is given by $r=0$,
and $\alpha=\mathrm{const.}$ on $\sigma_0$.
The extrinsic curvature of $S$ becomes
\begin{equation}
  \bar{K}_{ab}=-n_an_b\frac{{}^{(3)}\pounds_{\hat{r}}\alpha}{\alpha}+k_{ab}.
\end{equation}
Since $k_{ab}$ is symmetric, 
there exists an orthonormal basis
$(\mathbf{e}_1)_a$ and $(\mathbf{e}_2)_a$ that diagonalizes
$k_{ab}$ as
\begin{equation}
k_{ab} = k_{\rm 1}(\mathbf{e}_1)_a(\mathbf{e}_1)_b
+k_{\rm 2}(\mathbf{e}_2)_a(\mathbf{e}_2)_b,
\label{kab-diagonalized}
\end{equation}
and we introduce
\begin{subequations}
\begin{eqnarray}
  k_{\rm L} &=& \mathrm{max}(k_1,k_2), \label{def:k_L}\\
  k_{\rm S} &=& \mathrm{min}(k_1,k_2) \label{def:k_S}
\end{eqnarray}
\end{subequations}
for a later convenience. 
Then, the marginally transversely trapping condition
of Eq.~\eqref{marginally-transversely-trapping-condition},
is rewritten as 
\begin{equation}
  k_{\rm L} = \frac{{}^{(3)}\pounds_{\hat{r}}\alpha}{\alpha}.
  \label{marginally-transversely-trapping-condition-time-symmetric}
\end{equation}
By combining geometric equations, the following formula
for ${}^{(3)}\bar{\pounds}_{n}\bar{k}$ is derived \cite{Yoshino:2019}:
\begin{equation}
{}^{(3)}\bar{\pounds}_{n}\bar{k} = -\frac12{}^{(2)}R
-8\pi GP_r
+k\frac{{}^{(3)}\pounds_{\hat{r}}\alpha}{\alpha}
+\frac12\left(k^2-k_{ab}k^{ab}\right),
\label{Lie-derivative-bark-time-symmetric}
\end{equation}
where $P_{r}:=T_{ab}\hat{r}^a\hat{r}^b$ is the radial pressure,
and the Einstein field equations $G_{ab}=8\pi GT_{ab}$ have been assumed.
By substituting Eq.~\eqref{marginally-transversely-trapping-condition-time-symmetric}
into Eq.~\eqref{Lie-derivative-bark-time-symmetric} and
requiring ${}^{(3)}\bar{\pounds}_{n}\bar{k}$ to be zero,
we obtain the equation for marginally DTTSs: 
\begin{equation}
  -\frac12{}^{(2)}R -8\pi GP_{r} + k_1k_2+(k_1+k_2)\mathrm{max}(k_1,k_2)=0,
  \label{Equation-for-a-marginally-DTTS-time-symmetric}
\end{equation}
where the left-hand side
is expressed only in terms of geometrical and physical quantities on $\Sigma$. 
Since only vacuum initial data
are considered in this paper, the radial pressure is set to be zero,
$P_{r}=0$, in what follows.

%
%
\section{Setups and the equation for marginally DTTSs}
\label{section3}

In this section we describe the setups of the systems.
The three systems 
to be investigated are explained one by one in Sect.~\ref{section3-1}.
The methods of solving for marginally DTT spheres/tori
are briefly commented on in Sect.~\ref{section3-2}.

\subsection{Setups of the systems}
\label{section3-1}

In this paper we consider time-symmetric initial
data with conformally flat structure,
\begin{equation}
ds^2 = \varPsi^4(dx^2+dy^2+dz^2).
\end{equation}
The Hamiltonian constraint for a vacuum spacetime becomes
\begin{equation}
\bar{\nabla}^2\varPsi = 0,
\end{equation}
where $\bar{\nabla}^2$ denotes the flat-space Laplacian. 
We consider three kinds of solutions to this equation.

\subsubsection{Two-black-hole initial data}

The first one is a system of two black holes
momentarily at rest, 
\begin{equation}
  \varPsi = 1+\frac{GM}{4\sqrt{x^2+y^2+(z-z_0)^2}}
  +\frac{GM}{4\sqrt{x^2+y^2+(z+z_0)^2}}.
\end{equation}
This is called the Brill--Lindquist initial data \cite{Brill:1963}.
Although marginally DTTSs in this system
have been studied in our previous paper,
we examine the same system again 
from the viewpoint of the stretched hoop conjecture.

\subsubsection{Spindle initial data}

The second one is a system with a spindle source
located at $-L/2\le z\le L/2$ on the $z$-axis, 
\begin{eqnarray}
  \varPsi &=& 1+\frac{GM}{2L}\int_{-L/2}^{L/2}
  \frac{d\zeta}{\sqrt{x^2+y^2+(z-\zeta)^2}}. 
  \label{Psi-spindle-integral-form}
\end{eqnarray}
After integration, we obtain
\begin{equation}
\varPsi = 1+\frac{GM}{2L}\log\frac{r_+-z_+}{r_--z_-},
\end{equation}
with
\begin{subequations}
\begin{eqnarray}
  z_\pm &:=& z \mp {L}/{2},\\
  r_\pm &:=& \sqrt{x^2+y^2+z_\pm^2}.
\end{eqnarray}
\end{subequations}
We study this system in order to examine to what extent a
marginally DTTS can become long in one direction.
The same system was studied from the viewpoint
of the apparent horizon formation in Ref.~\cite{Chiba:1994}.

\subsubsection{Ring initial data}

The third one is a system with a ring-shaped source
located at a circle with radius $R$ on the $(x,y)$-plane, 
\begin{eqnarray}
  \varPsi &=& 1+\frac{GM}{4\pi} \int_{0}^{2\pi}
  \frac{d\zeta}{\sqrt{(x-R\cos\zeta)^2+(y-R\sin\zeta)^2+z^2}}.
  \label{Psi-ring-integral-form}
\end{eqnarray}
Using the complete elliptic integral of the first kind,
\begin{equation}
  K(\kappa) = \int_0^{\pi/2}\frac{d\zeta}{\sqrt{1-\kappa^2\sin^2\zeta}} 
  \qquad \textrm{with} \qquad \kappa := \sqrt{\frac{2b}{a+b}},
\end{equation}
the conformal factor is represented as 
\begin{equation}
\varPsi = 1+\frac{GM}{\pi\sqrt{a+b}}K(\kappa),
\label{Psi-ring-final-form}
\end{equation}
where $a$ and $b$ are defined by 
\begin{subequations}
\begin{eqnarray}
  a&:=& x^2+y^2+z^2+R^2,\\
  b&:=& 2R\sqrt{x^2+y^2}.
\end{eqnarray}
\end{subequations}
This system is chosen in order to examine
to what extent a marginally DTTS can become oblate.
Also, the formation of a marginally DTT/trapped torus is studied 
in this system.
The same system was studied from the viewpoint
of the apparent horizon formation in Ref.~\cite{Chiba:1994}.

\subsection{Method of solving for marginally DTTSs}
\label{section3-2}

We briefly describe how to solve for marginally DTT spheres
in the three systems and to solve for 
marginally DTT/trapped tori in the ring system.

\subsubsection{Marginally DTT spheres}

In order to solve for marginally DTT spheres,
it is convenient to introduce the spherical-polar
coordinates $(\tilde{r}, \tilde{\theta}, \phi)$ with
\begin{subequations}
\begin{eqnarray}
  x &=& \tilde{r}\sin\tilde{\theta}\cos\phi,\\
  y &=& \tilde{r}\sin\tilde{\theta}\sin\phi,\\
  z &=& \tilde{r}\cos\tilde{\theta}.
\end{eqnarray}
\end{subequations}
From the axial symmetry, the conformal factor $\varPsi$
depends only on $\tilde{r}$ and $\tilde{\theta}$.
We parametrize a surface $\sigma_0$ with the spherical topology as
\begin{equation}
  \tilde{r}=h(\tilde{\theta}).
  \label{sphere-parametrization}
\end{equation}
The equations for marginally DTT spheres
become second-order ordinary differential equations of $h(\tilde{\theta})$.
Those equations have been derived
in order to study the Brill--Lindquist two-black-hole
initial data in our previous paper \cite{Yoshino:2019},
and the same equations presented in terms of the conformal factor $\varPsi$
can be applied to spindle and
ring initial data as well. Hence, 
we refer interested readers to Ref.~\cite{Yoshino:2019}
for the explicit forms of the equations and their
derivation.
Those equations are solved under the boundary conditions
$h^\prime=0$ at $\tilde{\theta}=0$ and $\pi/2$.

\subsubsection{Marginally DTT/trapped tori}
\label{section3-2-2}

In order to study marginally DTT/trapped tori
in the ring system,
it is convenient to introduce the coordinates
$(\tilde{\rho}, \tilde{\xi}, \phi)$ by 
\begin{subequations}
\begin{eqnarray}
  x&=&(R+\tilde{\rho}\cos\tilde{\xi})\cos\phi,\\
  y&=&(R+\tilde{\rho}\cos\tilde{\xi})\sin\phi,\\
  z&=&\tilde{\rho}\sin\tilde{\xi}.
\end{eqnarray}
\end{subequations}
In these coordinates, the position of the ring
is given by $\tilde{\rho}=0$, and the metric becomes 
\begin{equation}
  ds^2 = \varPsi^4\left[d\tilde{\rho}^2 + \tilde{\rho}^2d\tilde{\xi}^2
    +(R+\tilde{\rho}\cos\tilde{\xi})^2d\phi^2\right].
\end{equation}
Parametrizing a toroidal surface $\sigma_0$ by
\begin{equation}
  \tilde{\rho}=h(\tilde{\xi}),
  \label{parametrization-torus}
\end{equation}
we perform the following coordinate transformations
from $(\tilde{\rho}, \tilde{\xi})$ to $(\rho, \xi)$:
\begin{eqnarray}
  \tilde{\rho} &=& \rho+h(\xi),
  \label{coordinate-transformation-tilderho-rho-xi}\\
  \tilde{\xi} &=& \xi-p(\rho,\xi),
  \label{coordinate-transformation-tildexi-rho-xi}
\end{eqnarray}
in order to calculate the orthonormal components of
the extrinsic curvature $k_{ab}$, i.e.  
$k_1$ and $k_2$ defined in Eq.~\eqref{kab-diagonalized}.
In the new coordinates, 
the surface $\sigma_0$ is given by $\rho=0$, and 
we require $\tilde{\xi}=\xi$ on the surface $\sigma_0$,
i.e. $p(0,\xi)=0$. Then, 
the orthonormal basis on $\sigma_0$ is introduced as
\begin{equation}
  \mathbf{e}_1=\varPsi^2\sqrt{h^{\prime 2}+h^2} \mathrm{d}\xi,
  \qquad
  \mathbf{e}_2=\varPsi^2(R+h\cos\xi) \mathrm{d}\phi.
  \label{orthonormal-basis-torus}
\end{equation}
The formulae for $k_1$ and $k_2$
and the equations for marginally DTT/trapped tori 
are presented in Appendix \ref{Appendix-A}.
Those equations are reduced to 
second-order ordinary differential equations for $h(\tilde{\xi})$,
and we numerically solve them under the boundary conditions
$h^\prime =0$ at $\tilde{\xi} = 0$ and $\pi$.

%
%
\section{Stretched hoop conjecture}
\label{section4}

In this section we study whether the inequality
$\mathcal{C}\lesssim 6\pi GM$ gives the condition
for the formation of outermost DTT spheres.
In Sect.~\ref{section4-1}, 
we briefly review the trials to formulate precisely the hoop conjecture
as previously discussed.
In Sect.~\ref{section4-2}, after commenting on how to
test the stretched hoop conjecture in this paper,
we present the numerical results. We summarize the
numerical results and discuss their implications in 
Sect.~\ref{section4-3}.

\subsection{Description of the hoop conjecture}
\label{section4-1}

The hoop conjecture is loosely formulated,
probably because its main purpose is to give an intuition
for the condition for horizon formation.
But when the hoop conjecture is tested, a more precise
formulation is required, and 
efforts in such a direction have been made in several
works \cite{Flanagan:1991,Chiba:1994}.\footnote{See
  also Refs.~\cite{Senovilla:2007,Gibbons:2009,Cvetic:2011,Gibbons:2012}
  for proposals
  on variants of the hoop conjecture.
 There is a debate on the proposal of Refs.~\cite{Gibbons:2009,Cvetic:2011,Gibbons:2012}: see Ref.~\cite{Mantoulidis:2014}.}
The ambiguous points are whether the horizon is an
apparent horizon or an event horizon, 
the definitions of the circumference $\mathcal{C}$
and the mass $M$, and the meaning of ``$\lesssim$'' (or,
in which situations the hoop conjecture is regarded to hold). 
As for the concept of the horizon, most works
adopt the apparent horizon, although there is also a study~\cite{Ida:1998}
that discussed the connection between event horizon formation
and the hoop conjecture.

In the formulation by Flanagan \cite{Flanagan:1991},
both the circumference and the mass are functions
of a closed surface $\sigma$, i.e. $\mathcal{C}(\sigma)$ and ${M}_{\rm Q}(\sigma)$,
and for each spacelike hypersurface the following quantity
(say, the hoop parameter) is determined:
\begin{equation}
  \mathcal{H}_{\rm A}
  = \mathrm{min}\left[\frac{\mathcal{C}(\sigma)}{4\pi GM_{\rm Q}(\sigma)}\right].
\end{equation}
Several candidates for the definition $\mathcal{C}(\sigma)$ are discussed
in Refs.~\cite{Flanagan:1991,Chiba:1994}, and there does
not seem to be a consensus.
As for an axisymmetric surface $\sigma$, the following definition
seems to be widely accepted:
\begin{equation}
  \mathcal{C}(\sigma) = \mathrm{max}(\mathcal{C}_{\rm p},\ \mathcal{C}_{\rm e}),
  \label{Circumference-axisymmetric-case}
\end{equation}
where $\mathcal{C}_{\rm p}$ is the polar circumference that is twice
the proper distance between the north and south poles,
and $\mathcal{C}_{\rm e}$ is the maximum length of closed azimuthal curves.
See also Ref.~\cite{Chiba:1998} for the application of the
definition of $\mathcal{C}(\sigma)$
for non-axisymmetric surfaces proposed in Ref.~\cite{Chiba:1994}.

The definition of the mass is also an open problem.
Several negative arguments against the hoop conjecture 
were made by studying a static charged star \cite{Bonnor:1983,Bonnor:1984,Leon:1987},
but it was pointed out that the evaluation of the mass was not
appropriate because the energy of electric fields distributes
outside of the surface on which the circumference is
evaluated \cite{Flanagan:1991,Hod:2018,Peng:2019}.  
The problem is that 
local gravitational mass cannot be determined uniquely
in general relativity,
and there are many candidates for the ``quasilocal 
mass'' $M_{\rm Q}(\sigma)$ associated with a surface $\sigma$. 
In Refs.~\cite{Bernstein:1994,Yoshino:2001},
the hoop conjecture was tested using 
Penrose's quasilocal mass \cite{Penrose:1982}
and Hawking's quasilocal mass \cite{Hawking:1968}, respectively,
and it was reported that the hoop conjecture holds better
if the quasilocal definitions of 
masses are used rather than the ADM mass.\footnote{
  See also Refs.~\cite{Murchadha:2009,Malec:2015} for
  studies on the hoop conjecture
  with the Brown--York quasilocal mass, $M_{\rm BY}$ \cite{Brown:1992}.
  In this case, the hoop conjecture takes the form
  $\mathcal{C}\lesssim 2\pi GM_{\rm BY}$
  due to the property of the Brown--York mass.}
In systems where the amount of energy of matter or junk gravitational radiation
outside $\sigma$ is small, the hoop conjecture holds well
with the ADM mass \cite{Nakamura:1988,Chiba:1994,Chiba:1998}.

The criterion for the statement that the hoop conjecture holds
(or does not hold) is also not very clear.
The natural criterion would be as follows.
Consider a collection of spacelike hypersurfaces, which may be
a sequence of time evolution, or may be a set of initial data.
If there are two values $\mathcal{H}_{\rm A}^{\rm (S)}$ and
$\mathcal{H}_{\rm A}^{\rm (N)}$ such that
the apparent horizon is present 
if $\mathcal{H}_{\rm A}\le \mathcal{H}_{\rm A}^{\rm (S)}$
and is absent if $\mathcal{H}_{\rm A}\ge \mathcal{H}_{\rm A}^{\rm (N)}$,
the hoop conjecture is regarded to hold for that collection
of hypersurfaces,
because both the necessary and sufficient
conditions for the apparent horizon formation
are given in terms of $\mathcal{H}_{\rm A}$.
Although this criterion was explicitly stated in Ref.~\cite{Yoshino:2001}
for the first time, 
the existing works seem to have adopted this criterion implicitly.

\subsection{Examination of the stretched hoop conjecture}
\label{section4-2}

We now turn our attention to the stretched hoop conjecture
for the formation of outermost DTT spheres. 
Since we would like to examine whether $\mathcal{C}\lesssim 6\pi GM$
gives the condition for the formation of DTTSs, we consider
\begin{equation}
  \mathcal{H}_{\rm D} = \mathrm{min}
  \left[\frac{\mathcal{C}(\sigma)}{6\pi GM_{\rm Q}(\sigma)}\right]
\end{equation}
as the stretched hoop parameter to be studied. 
In this paper we adopt the ADM mass as the definition of mass,
that is, $M_{\rm Q}(\sigma)=M$ for arbitrary surfaces $\sigma$, because
the initial data are vacuum and no energy density of matter is present
outside $\sigma$. 
Furthermore, since the initial data are
time symmetric and conformally flat,
junk energy of gravitational waves is expected to be small.
Then, the problem is reduced to finding the minimum value of $\mathcal{C}(\sigma)$.

By virtue of the axial symmetry of the systems,
we adopt Eq.~\eqref{Circumference-axisymmetric-case}
as the definition of the circumference.
For prolate systems (i.e. the two-black-hole initial data and
the spindle initial data), on the one hand, 
the minimum value of $\mathcal{C}(\sigma)$ coincides with
the minimum value of the polar circumference $\mathcal{C}_{\rm p}(\sigma)$. 
Parametrizing the surface $\sigma$
in the same manner as Eq.~\eqref{sphere-parametrization},
the value of $\mathcal{C}_{\rm p}$ is calculated from
\begin{equation}
\mathcal{C}_{\rm p} = \int_0^{\pi}\varPsi^2 \sqrt{h^2+h^{\prime 2}}d\tilde{\theta}.
\end{equation}
From the variational principle, the equation for $h(\tilde{\theta})$
is derived as
\begin{equation}
  h^{\prime\prime}-2\left(\frac{\varPsi_{,\tilde{r}}}{\varPsi}+\frac{1}{h}\right)h^{\prime 2}
  -\left(2\frac{\varPsi_{,\tilde{r}}}{\varPsi}+\frac{1}{h}\right)h^2
  +2\frac{\varPsi_{,\tilde{\theta}}}{\varPsi}
  \left(1+\frac{h^{\prime 2}}{h^2}\right)h^{\prime}
  =0.
  \label{Equation-Hoop-prolate}
\end{equation}
The same equation is presented in Refs.~\cite{Nakamura:1988,Chiba:1994}. 
For oblate systems (i.e. the ring initial data), on the other hand, 
the minimum value of $\mathcal{C}(\sigma)$ coincides with
the minimum circumference of circles $\tilde{r}=\mathrm{const.}$
on the equatorial plane:
\begin{equation}
\mathcal{C}_{\rm e} = \mathrm{min} \left[2\pi \tilde{r}\varPsi^2(\tilde{r},\pi/2)\right].
\end{equation}
The value of $\tilde{r}$ at which $\mathcal{C}_{\rm e}$ becomes
minimal is determined by the equation
$\varPsi+2\tilde{r}\varPsi_{,\tilde{r}}=0$ on the equatorial plane. 
Below, we present the numerical results
for the three systems, one by one. 

\subsubsection{Two-black-hole initial data}

%
\begin{figure}[tb]
\centering
\includegraphics[width=0.43\textwidth,bb=0 0 297 311]{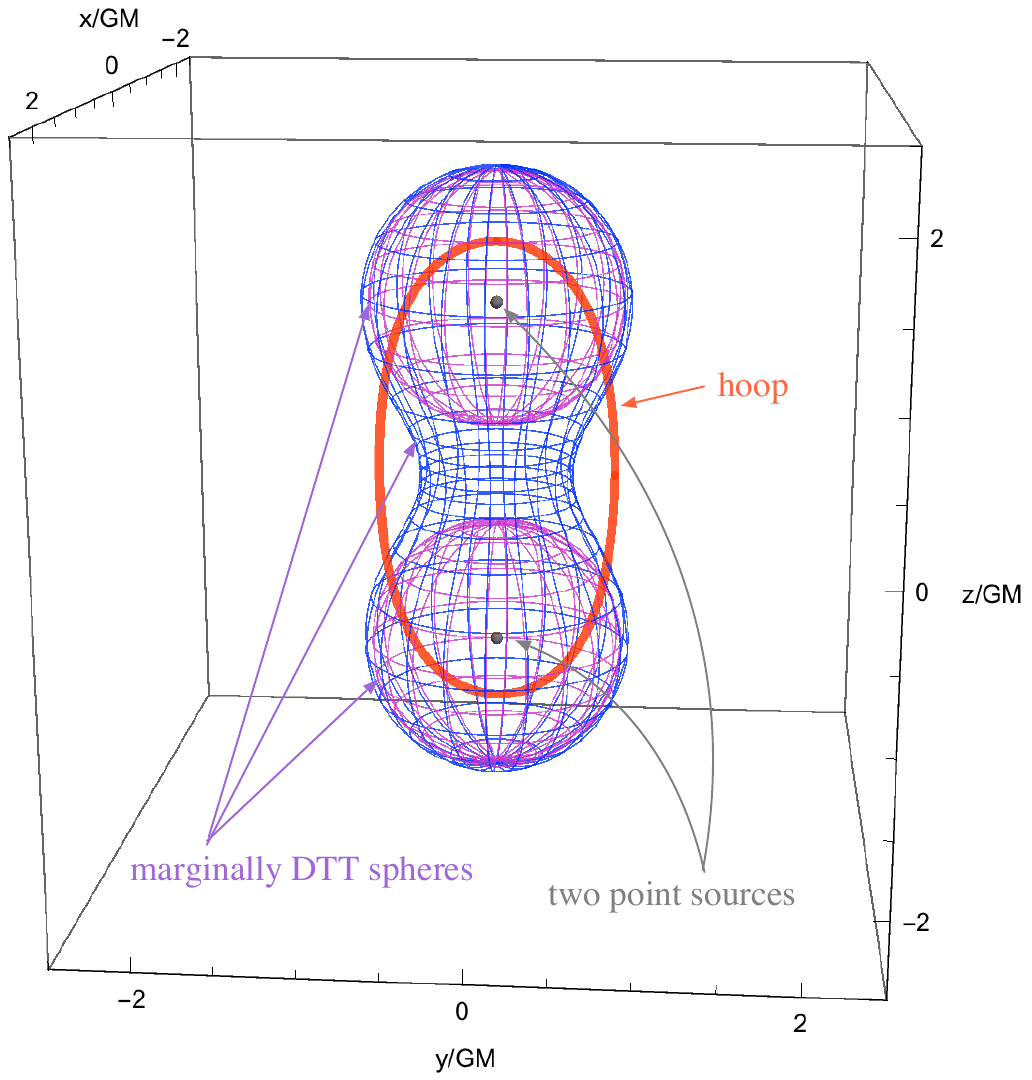}
\caption{THree dimensional plot of the marginally DTT spheres for
  $z_0/GM=1.1506$ in the two-black-hole initial data.
  The shortest hoop that surrounds the system
  is also shown.
}
\label{3D-plot-2BH}
\end{figure}
%

%
\begin{figure}[tb]
\centering
\includegraphics[width=0.5\textwidth,bb=0 0 414 299]{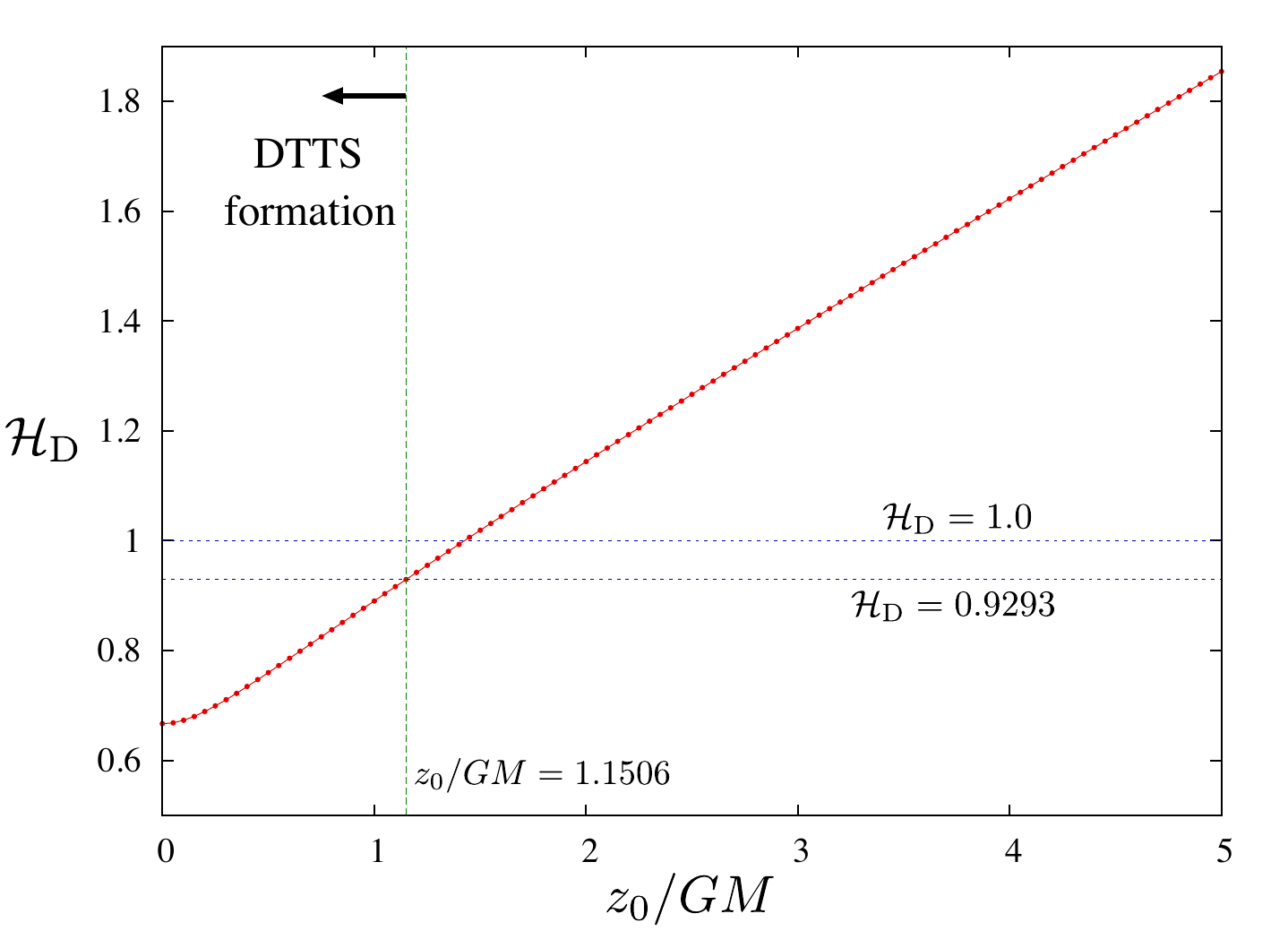}
\caption{The stretched hoop parameter $\mathcal{H}_{\rm D}$
  as a function of $z_0/GM$ in the initial data of two black holes.
  The value $\mathcal{H}_{\rm D}=0.9293$ is indicated by a horizontal dashed
  line, and the marginally DTT sphere forms if $\mathcal{H}_{\rm D}\le 0.9293$
  holds. The value $\mathcal{H}_{\rm D}=1.0$ is also indicated for
  comparison. }
\label{figure_hoop_2BH_edited}
\end{figure}
%

The marginally DTT spheres in two-black-hole systems
have been calculated in our previous
paper \cite{Yoshino:2019}, and 
the (common) marginally DTT sphere that surrounds both black holes
exists for $z_0/GM\lesssim 1.1506$. A three-dimensional (3D) plot
of the marginally DTT spheres for $z_0/GM=1.1506$
is shown in Fig.~\ref{3D-plot-2BH}, 
together with the shortest hoop that surrounds both
black holes determined from Eq.~\eqref{Equation-Hoop-prolate}.
Figure~\ref{figure_hoop_2BH_edited} shows the behavior
of the stretched hoop parameter $\mathcal{H}_{\rm D}$ as a function
of $z_0/GM$. The stretched hoop parameter $\mathcal{H}_{\rm D}$
is a monotonically increasing function of $z_0/GM$,
and if 
the relation $\mathcal{H}_{\rm D}\le 0.9293$
(respectively, $\mathcal{H}_{\rm D}\ge 0.9294$) holds,
the (common) marginally DTT sphere is present (respectively, absent).

\subsubsection{Spindle initial data}

%
\begin{figure}[tb]
\centering
\includegraphics[width=0.9\textwidth,bb=0 0 314 166]{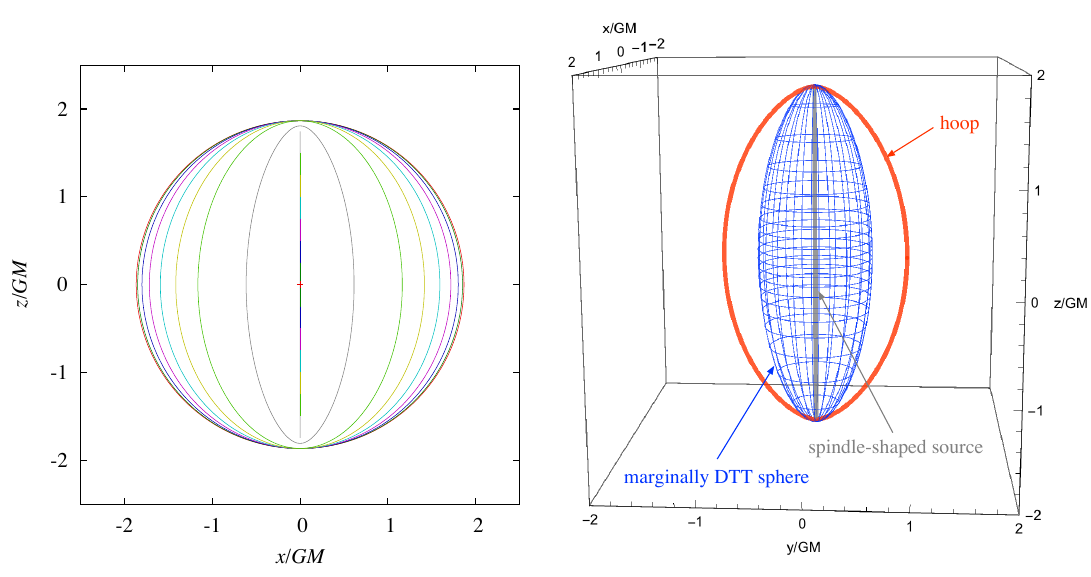}
\caption{Marginally DTT spheres in the spindle initial data.
  Left panel: Sections of marginally DTT spheres 
  with the $(x,z)$-plane for $L/GM=0.0$, $0.5$, $1.0$,
  $1.5$, $2.0$, $2.5$, $3.0$ and $3.4928$ in spindle systems.
  For $R/GM\ge 3.4929$, a marginally DTT sphere cannot be found.
  Right panel: 3D plot of the marginally DTT sphere for
  $L/GM=3.4928$. The shortest hoop that surrounds the system
  is also shown.
}
\label{figure_spindle_DTTS_all}
\end{figure}
%

%
\begin{figure}[tb]
\centering
\includegraphics[width=0.5\textwidth,bb=0 0 412 302]{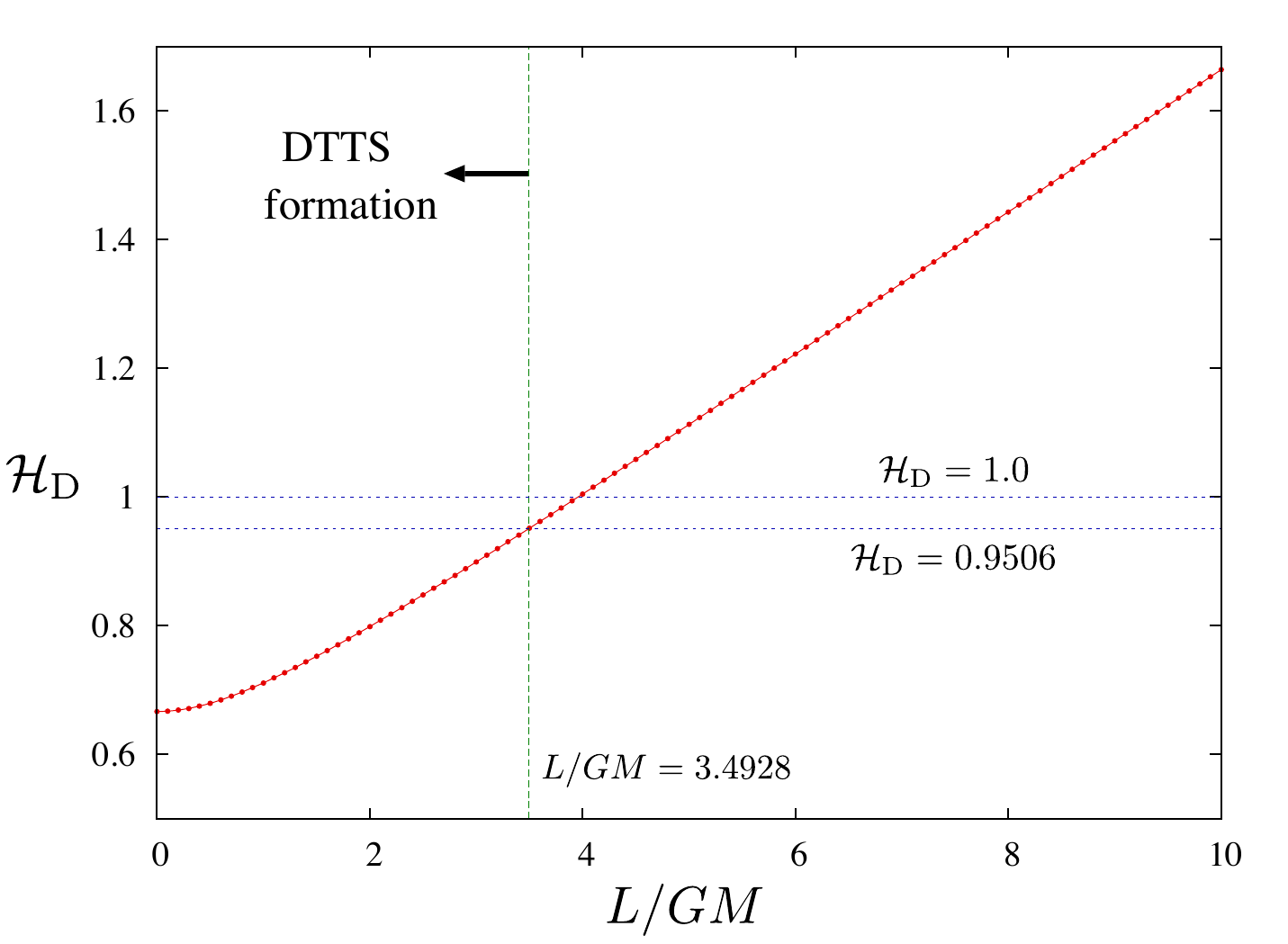}
\caption{The stretched hoop parameter $\mathcal{H}_{\rm D}$
  as a function of $L/GM$ in the spindle initial data.
  The value $\mathcal{H}_{\rm D}=0.9506$ is indicated by a horizontal dashed
  line, and the marginally DTT sphere forms if $\mathcal{H}_{\rm D}\le 0.9506$
  holds. The value $\mathcal{H}_{\rm D}=1.0$ is also indicated for
  comparison. 
}
\label{figure_hoop_spindle_edited}
\end{figure}
%

The left panel of Fig.~\ref{figure_spindle_DTTS_all} depicts
marginally DTT spheres for
values of $L/GM$ from $0.0$ to $3.0$ at $0.5$ intervals,
and for $3.4928$.
As the value of $L/GM$ is increased, the marginally DTT sphere
becomes more prolate. 
For $L/GM= 3.4928$, the marginally DTT sphere 
approximately degenerates 
with the inner boundary of the dynamically transversely trapping region
(that is not plotted in Fig.~\ref{figure_spindle_DTTS_all}),
and they vanish for $L/GM\ge 3.4929$.
Therefore, a marginally DTT sphere cannot become arbitrarily long
in the $z$ direction. This result is similar to the
apparent horizon formation in the same initial data
studied in Ref.~\cite{Chiba:1994}, while it is 
in contrast to the apparent horizon formation 
in the higher-dimensional version of
these initial data where the apparent horizon
can become arbitrarily long \cite{Ida:2002}.
The right panel of Fig.~\ref{figure_spindle_DTTS_all}
shows a 3D plot of the marginally DTT sphere for
$L/GM=3.4928$ together with the shortest hoop that surrounds
the system.

Figure~\ref{figure_hoop_spindle_edited} shows the behavior
of the stretched hoop parameter $\mathcal{H}_{\rm D}$ as a function
of $L/GM$. The stretched hoop parameter $\mathcal{H}_{\rm D}$
is a monotonically increasing function of $L/GM$,
and if 
the relation $\mathcal{H}_{\rm D}\le 0.9506$
(respectively, $\mathcal{H}_{\rm D}\ge 0.9507$) holds,
the marginally DTT sphere is present (respectively, absent).

\subsubsection{Ring initial data}

%
\begin{figure}[tb]
\centering
\includegraphics[width=0.9\textwidth,bb=0 0 395 196]{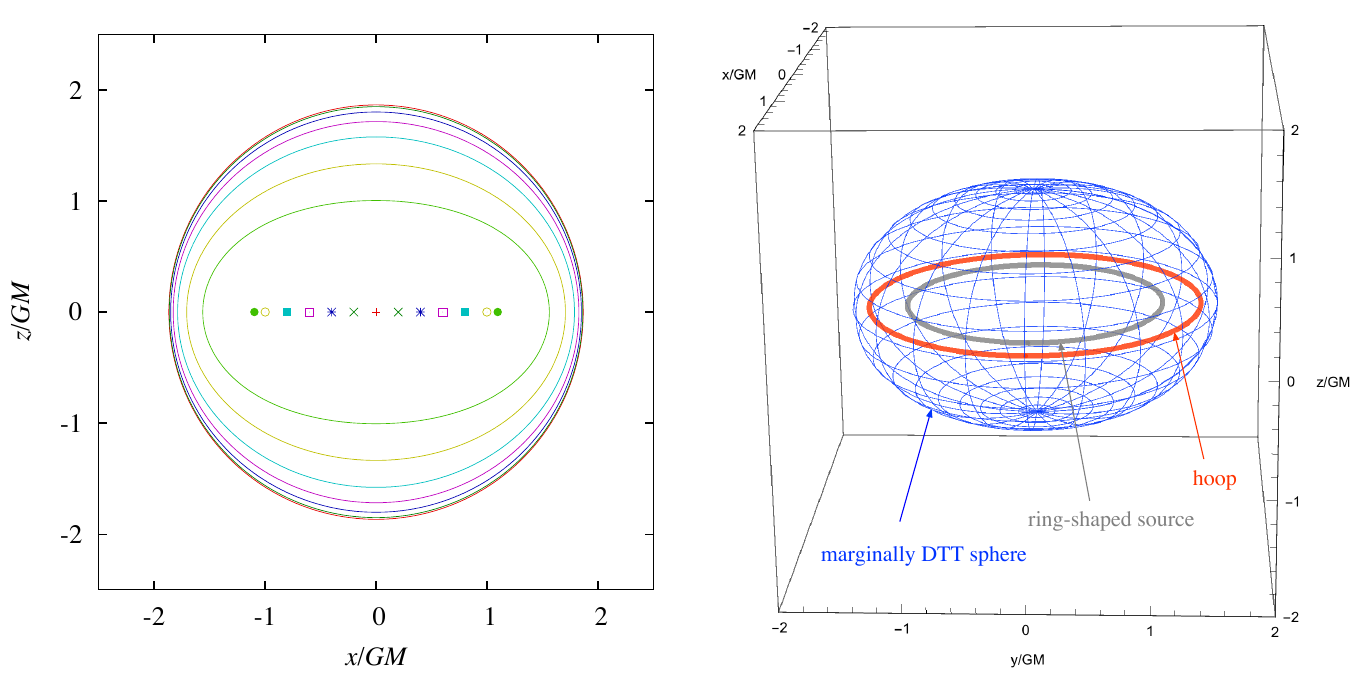}
\caption{Marginally DTT spheres in the ring initial data.
  Left panel: Sections of marginally DTT spheres 
  with the $(x,z)$-plane for $R/GM=0.0$, $0.2$, $0.4$,
  $0.6$, $0.8$, $1.0$, and $1.0943$ in ring systems.
  For $R/GM\ge 1.0944$, a marginally DTT sphere cannot be found.
  Right panel: 3D plot of the marginally DTT sphere for
  $R/GM=1.0943$. The shortest hoop that surrounds the system
  is also shown.
}
\label{figure_ring_DTTS_all}
\end{figure}
%

%
\begin{figure}[tb]
\centering
\includegraphics[width=0.5\textwidth,bb=0 0 417 302]{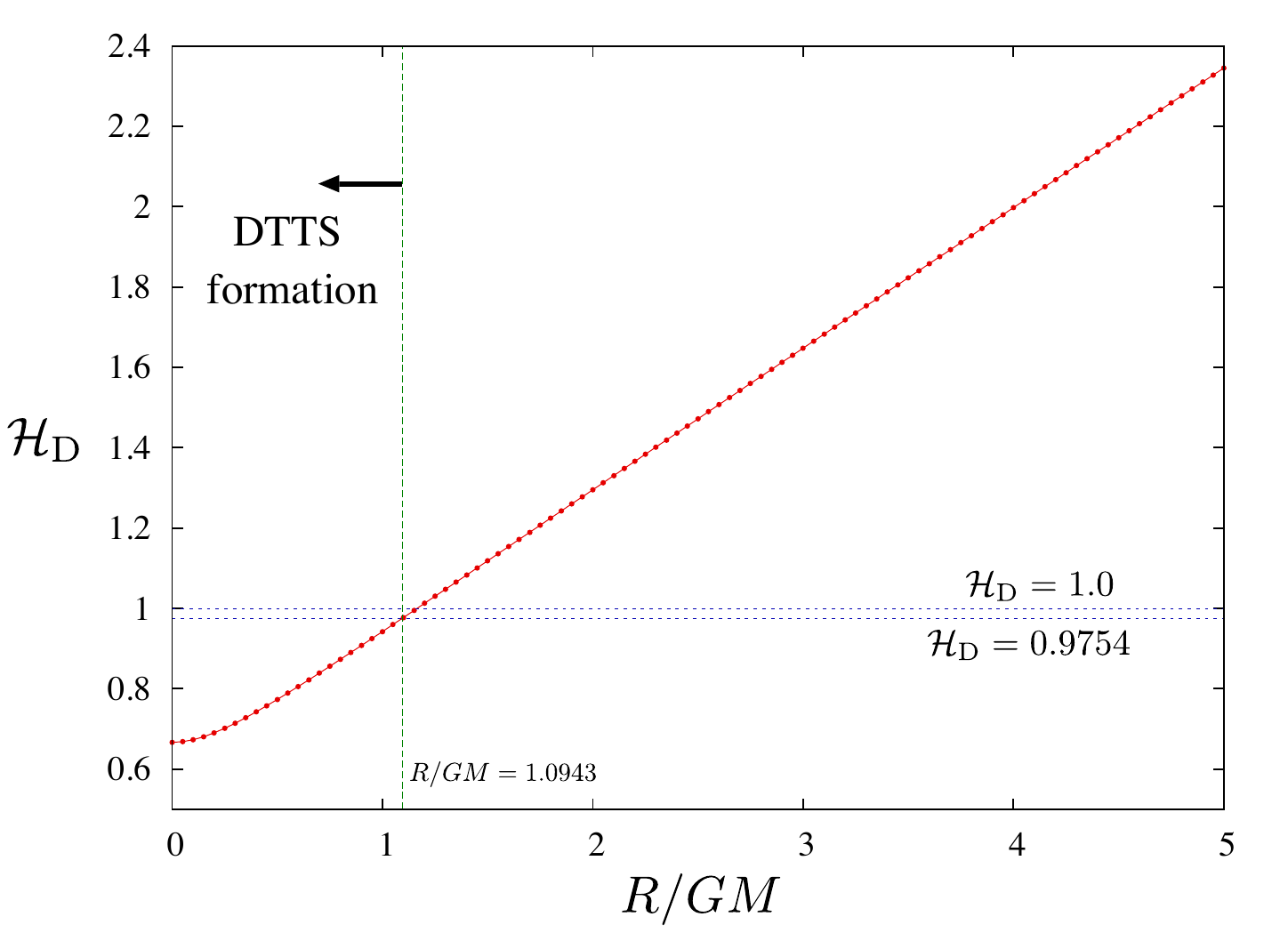}
\caption{The stretched hoop parameter $\mathcal{H}_{\rm D}$
  as a function of $R/GM$ in the ring initial data.
  The value $\mathcal{H}_{\rm D}=0.9754$ is indicated by a horizontal dashed
  line, and the marginally DTT sphere forms if $\mathcal{H}_{\rm D}\le 0.9754$
  holds. The value $\mathcal{H}_{\rm D}=1.0$ is also indicated for
  comparison. 
}
\label{figure_hoop_ring_edited}
\end{figure}
%

The left panel of Fig.~\ref{figure_ring_DTTS_all} depicts
marginally DTT spheres for
values of $R/GM$ from $0.0$ to $1.0$ at $0.2$ intervals,
and for $1.0943$.
As the value of $R/GM$ is increased, the marginally DTT sphere
becomes more oblate. 
For $R/GM= 1.0943$, the marginally DTT sphere 
approximately degenerates 
with the inner boundary of the dynamically transversely trapping region
(that is not plotted in Fig.~\ref{figure_ring_DTTS_all}),
and they vanish for $R/GM\ge 1.0944$.
The right panel of Fig.~\ref{figure_ring_DTTS_all} shows a
3D plot of the marginally DTT sphere for $R/GM=1.0943$ together
with the shortest hoop that surrounds the system.

Figure~\ref{figure_hoop_ring_edited} shows the behavior
of the stretched hoop parameter $\mathcal{H}_{\rm D}$ as a function
of $R/GM$. The stretched hoop parameter $\mathcal{H}_{\rm D}$
is a monotonically increasing function of $R/GM$,
and if 
the relation $\mathcal{H}_{\rm D}\le 0.9754$
(respectively, $\mathcal{H}_{\rm D}\ge 0.9755$) holds,
the marginally DTT sphere is present (respectively, absent).

\subsection{Summary of the numerical results}
\label{section4-3}

In each of the three systems,
the region where the DTT sphere is present/absent
can be specified by 
the values of $\mathcal{H}_{\rm D}$.
Putting the three results together,
a marginally DTT sphere forms if $\mathcal{H}_{\rm D}\le 0.9294$,
and it does not form if $\mathcal{H}_{\rm D}\ge 0.9755$.
Therefore, the stretched hoop parameter $\mathcal{H}_{\rm D}$
becomes an indicator for the formation of a marginally DTT sphere
at least in the systems studied in this paper.
From this result, 
it would be fair to present the following stretched hoop conjecture:

%
\begin{conjecture}
Strong gravity regions with marginally DTT spheres form
when and only when a mass $M$
gets compacted into a region whose circumference in every direction
is bounded by $\mathcal{C}\lesssim 6\pi GM$.
\end{conjecture}
%

The stretched hoop conjecture suggests that
an arbitrarily long DTT sphere cannot form.
As discussed in Sect.~\ref{section1}, 
this also indicates that a marginally DTT torus
could not form, or even if it does form, it would be located inside
a marginally DTT sphere.
In the next section, we investigate the ring system further,
and show that there is a parameter region of $R/GM$
where a marginally TTS torus actually forms
inside a marginally TTS sphere.

%
%
\section{Further examination of the ring system}
\label{section5}

As mentioned in Sect.~\ref{section1}, there
is the possibility that DTT tori may form
unless they are convex [more strictly, unless $k_{\rm S}\ge -k_{\rm L}/3$
is satisfied, where $k_{\rm S}$ and
$k_{\rm L}$ are defined in Eqs.~\eqref{def:k_L} and \eqref{def:k_S}].
In Sect.~\ref{section5-1},
we will show that in a ring system there is a parameter
region where a marginally DTT torus is actually present
inside a marginally DTT sphere.
In Sect.~\ref{section5-2}, we present the similarity
between the two concepts of a DTTS and a trapped surface by showing that
there are configurations where a marginally trapped torus
forms inside a marginally trapped sphere in the same system.

\subsection{Marginally DTT tori}
\label{section5-1}

%
\begin{figure}[tb]
\centering
\includegraphics[width=0.7\textwidth,bb=0 0 360 192]{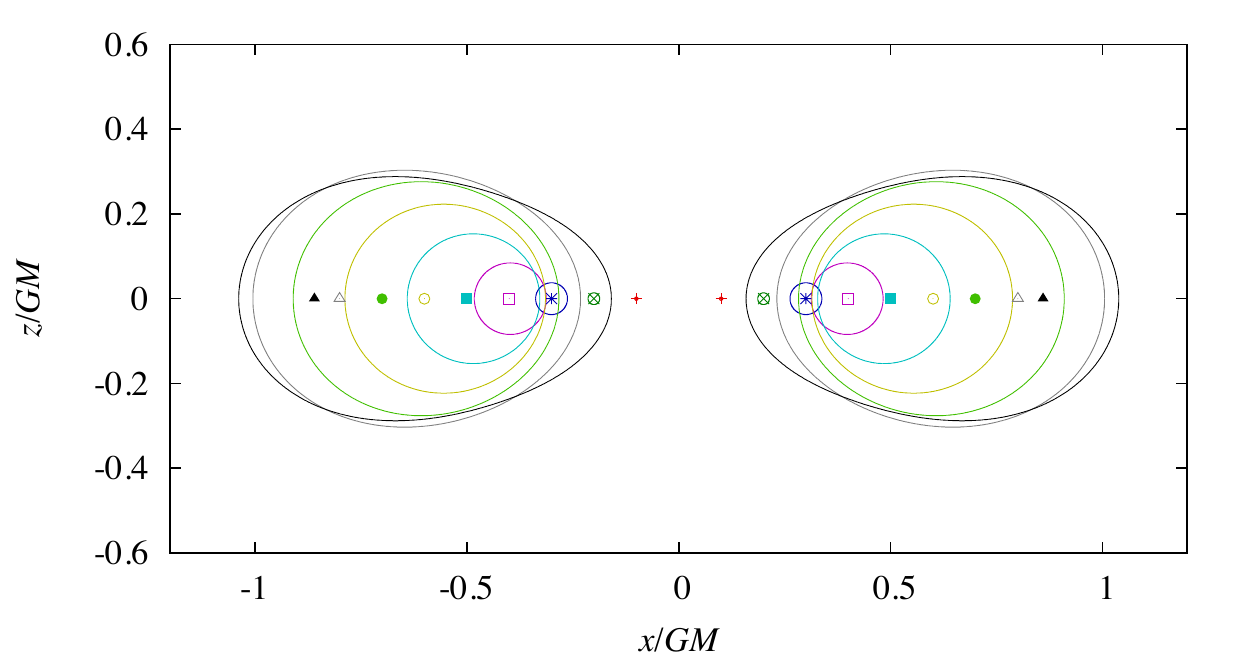}
\caption{Sections of marginally DTT tori in ring systems 
  with the $(x,z)$-plane for $R/GM=0.1$, $0.2$, $0.3$,
  $0.4$, $0.5$, $0.6$, $0.7$, $0.8$, and $0.8598$.
  For $R/GM\ge 0.8599$, a marginal DTT torus cannot be found.
}
\label{figure_ring_torus_DTTS_all}
\end{figure}
%

%
\begin{figure}[tb]
\centering
\includegraphics[width=0.9\textwidth,bb=0 0 459 182]{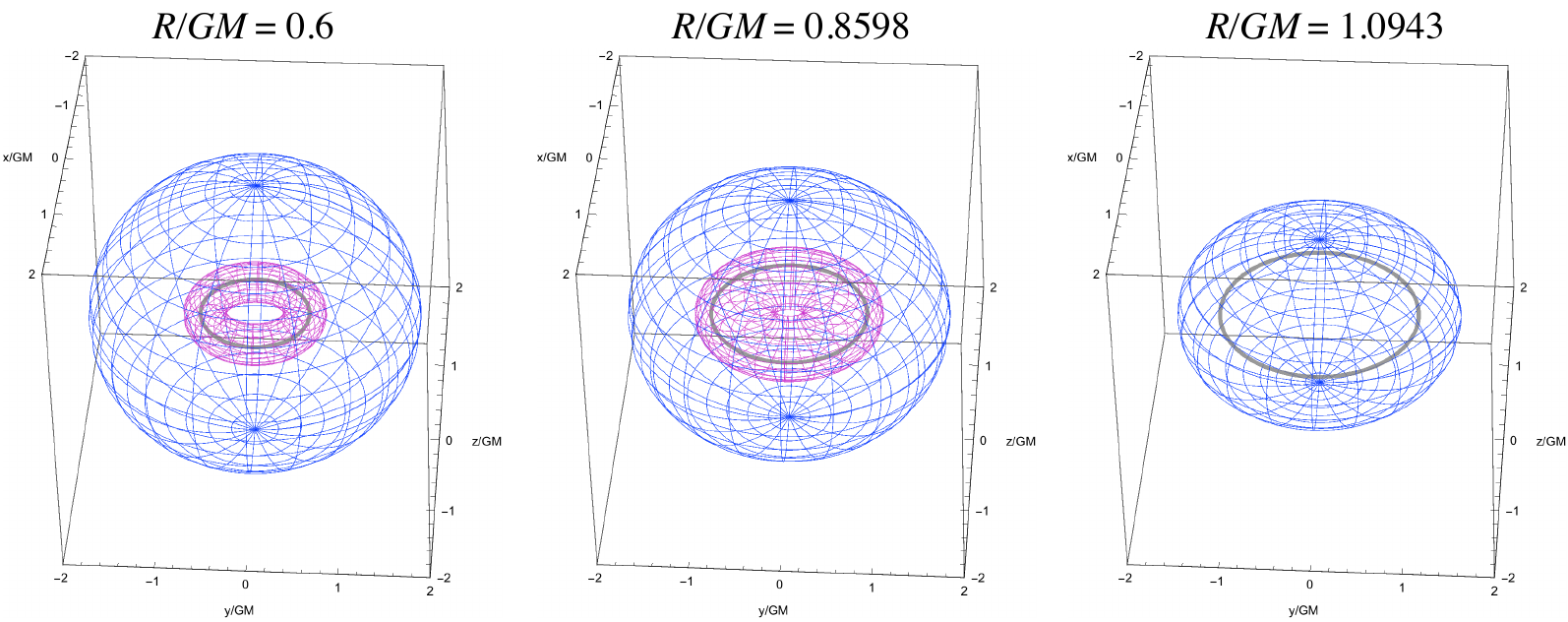}
\caption{Three-dimensional plots
  of a marginally DTT sphere and a marginally DTT torus
  for $R/GM=0.6$ (left panel),
  $0.8598$ (middle panel), and $1.0943$ (right panel).
}
\label{figure_torus_spherical_TTS_radius}
\end{figure}
%

As a result of numerical survey, we have found that
a marginally DTT torus actually exists in the parameter
range $0<R/GM\le 0.8598$. 
Figure~\ref{figure_ring_torus_DTTS_all} shows sections of a marginally
DTT torus in the $(x, z)$-plane
for values of $R/GM$ from $0.1$ to $0.8$ at $0.1$ intervals,
and for $0.8598$.
No solution has been found for $R/GM\ge 0.8599$. Since
a marginally DTT sphere exists in the range $R/GM\le 1.0943$,
a marginally DTT torus is located inside a marginally DTT sphere.
In Fig.~\ref{figure_torus_spherical_TTS_radius}, a marginally
DTT torus and a marginally DTT sphere are plotted together
for $R/GM = 0.6$ (left panel),
$0.8598$ (middle panel), and $1.0943$ (right panel).

%
\begin{figure}[tb]
\centering
\includegraphics[width=0.55\textwidth,bb=0 0 360 252]{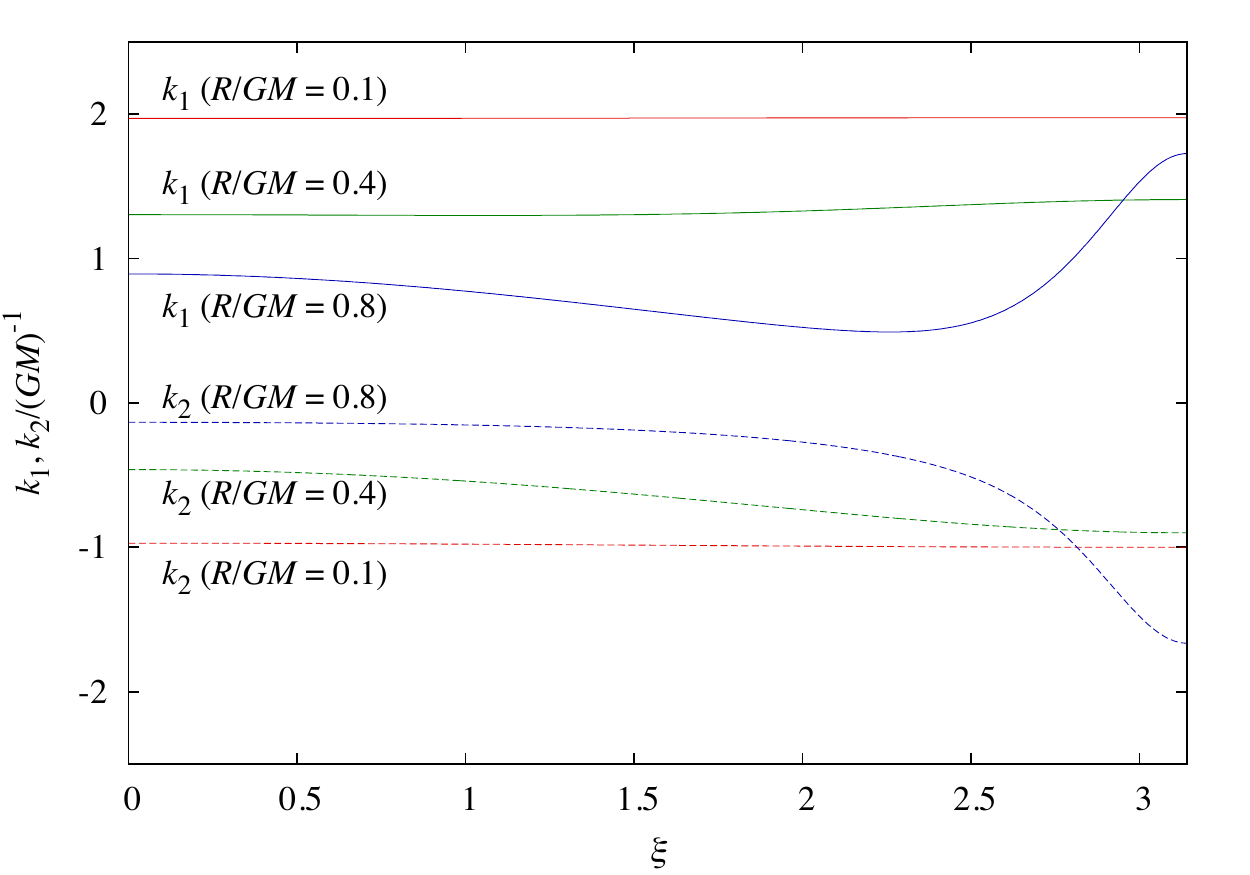}
\caption{The values of $k_1$ (solid lines) and $k_2$ (dashed lines)
  of marginally DTT tori 
  as functions of the angular coordinate $\tilde{\xi}$
  for $R/GM=0.1$, $0.4$, and $0.8$. 
}
\label{figure_torus_k1k2}
\end{figure}
%

Since it has been proved that a DTTS must have spherical topology
as long as $k_{\rm S}\ge -k_{\rm L}/3$ \cite{Yoshino:2019},
the obtained DTT tori must violate this inequality.
Figure~\ref{figure_torus_k1k2} plots the value of $k_1$ (solid curves)
and $k_2$ (dashed curves)
as functions of the angular coordinate $\tilde{\xi}$
for $R/GM=0.1$, $0.4$, and $0.8$. For all values of $\tilde{\xi}$,
the relation $k_2<0<k_1$ is maintained, and thus the marginally DTT tori
are not convex. Also, at least in the neighborhood of $\tilde{\xi}=\pi$,
the relation $k_2\ge -k_1/3$ is violated.
For $R/GM\ll 1$, the relation $k_1\approx -2k_2$ holds approximately.

The validity of these numerical results is checked as follows.
Integrating Eq.~\eqref{Equation-for-a-marginally-DTTS-time-symmetric}
on a marginally DTTS,
we have
\begin{equation}
\Delta :=\int_{\sigma_0} k_1(k_1+2k_2)dA-2\pi\chi(\sigma_0) =0,
\end{equation}
where $\chi(\sigma_0)$ is the Euler characteristic, which
is zero for a toroidal surface.
The numerical value of $\Delta$
deviates from zero due to numerical errors, and 
the error decreases as the numerical accuracy is systematically
increased if the calculation is correct.
By contrast, if there is a mistake somewhere,
the value of $\Delta$ does not decrease by this procedure.
By increasing the number of grid points up to  
$10^4$ and increasing the accuracy
of the initial condition in the shooting method
up to $O(10^{-12})$, the typical value of $\Delta$ is decreased
to the order of $10^{-10}$. This result supports the correctness
of our numerical calculation. Also, as demonstrated
in Appendix~\ref{Appendix-B},
an approximate solution of a marginally DTT torus
can be obtained for a small ring with $R\ll GM$, and it is
also consistent with our numerical results.

%
\begin{figure}[tb]
\centering
\includegraphics[width=0.55\textwidth,bb=0 0 407 286]{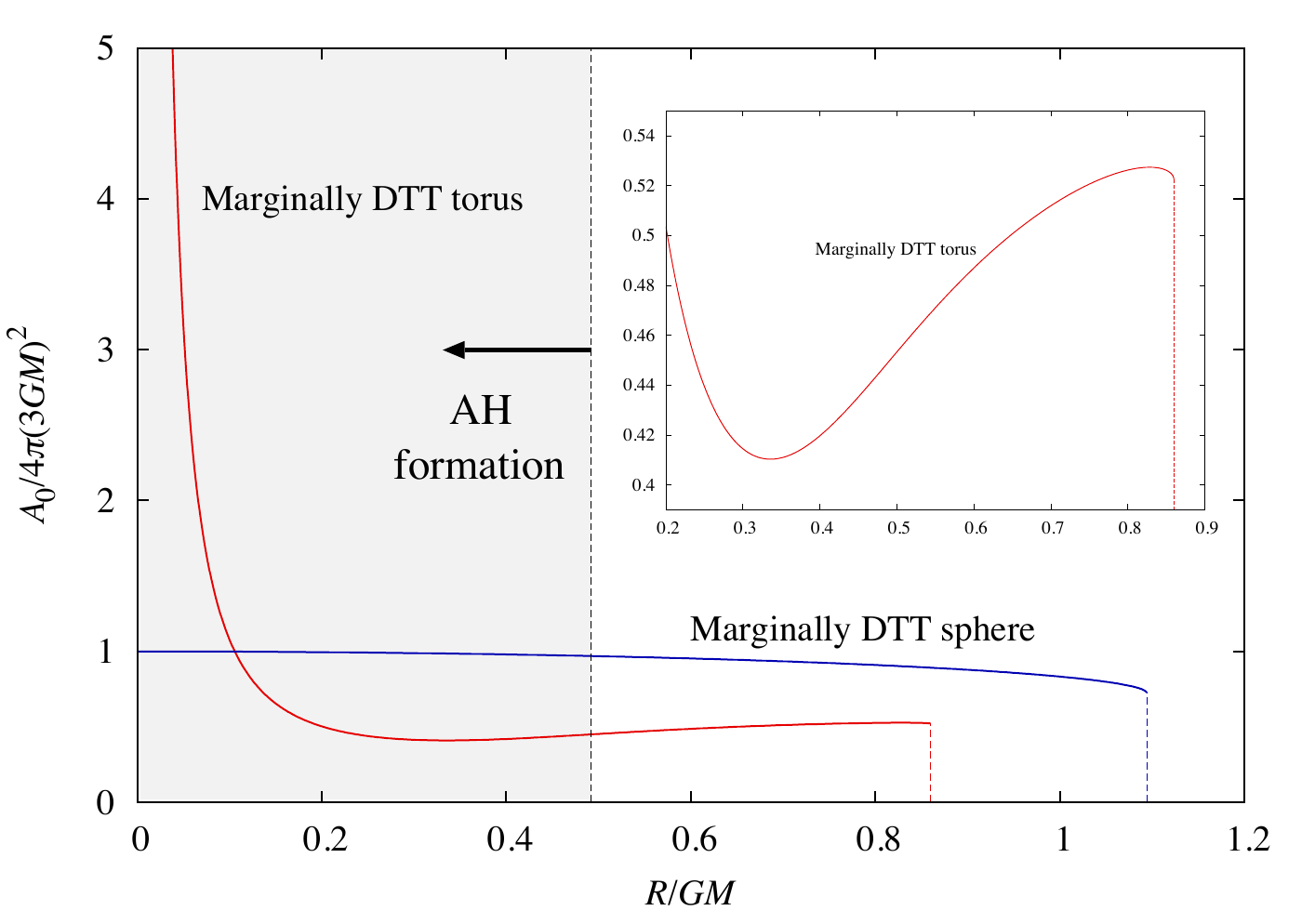}
\caption{The area $A_{0}$ of a marginally DTT sphere and
  that of a marginally DTT torus
  as functions of $R/GM$.
  The area is 
  normalized by $4\pi (3GM)^2$. 
  The parameter region $0\le R/GM\le 0.49203$
  where an apparent horizon
  forms is indicated by shaded regions. The inset enlarges
  the behavior of $A_{0}/4\pi (3GM)^2$ for a marginally DTT torus
  in the domain $0.2\le R/GM\le 0.9$.
}
\label{figure_area_ring_combine}
\end{figure}
%

Figure~\ref{figure_area_ring_combine} shows the area of the marginally DTT
sphere and the marginally DTT torus normalized by the area of
a photon sphere in a spherically symmetric case, $A_0/4\pi(3GM)^2$,
as functions of $R/GM$.
The area of the marginally DTT sphere is 
a monotonically decreasing function of $R/GM$, and takes
the value $A_0/4\pi(3GM)^2\approx 0.7274$ at $R/GM=1.0943$.
Therefore, the area satisfies the Penrose-like inequality,
$A_0\le 4\pi (3GM)^2$.
By contrast, the area of the marginally DTT torus does not
show monotonic behavior. Also, in the limit $R/GM\to 0$,
the area becomes indefinitely large. This is because
the condition $k_{\rm S}\ge -k_{\rm L}/3$ is not satisfied
on a marginally DTT torus, and hence 
the Penrose-like inequality proved in our previous paper
\cite{Yoshino:2019} does not apply to it.
Unfortunately, such a large DTT torus may not have important
physical meaning because
it is hidden inside an apparent horizon that forms  
for $R/GM\lesssim 0.49203$.
However, we would like to point out that there is a parameter region
of $R/GM$ where a marginally DTT torus exists without being hidden
by an apparent horizon.
Although we are not sure whether 
an event horizon that encloses marginally DTT torus
exists or not 
in the present analysis of initial data, 
this result indicates that
it may be possible to observe the positions
where marginally DTT tori exist.

\subsection{Marginally trapped tori}
\label{section5-2}

In our previous paper \cite{Yoshino:2019} we compared
marginally DTTSs and marginally trapped surfaces
for the Brill--Lindquist initial data of two equal-mass
black holes, and stressed the similarity between them.
Here, we would like to proceed with similar analysis
of the ring system. In particular, there are configurations
where a marginally trapped torus forms inside a marginally
trapped sphere. 
Although the formation of trapped tori
has been reported in other systems \cite{Husa:1996,Karkowski:2017},
the trapped torus formation in the present system 
is reported for the first time, to the best of our knowledge.

%
\begin{figure}[tb]
\centering
\includegraphics[width=0.45\textwidth,bb=0 0 260 252]{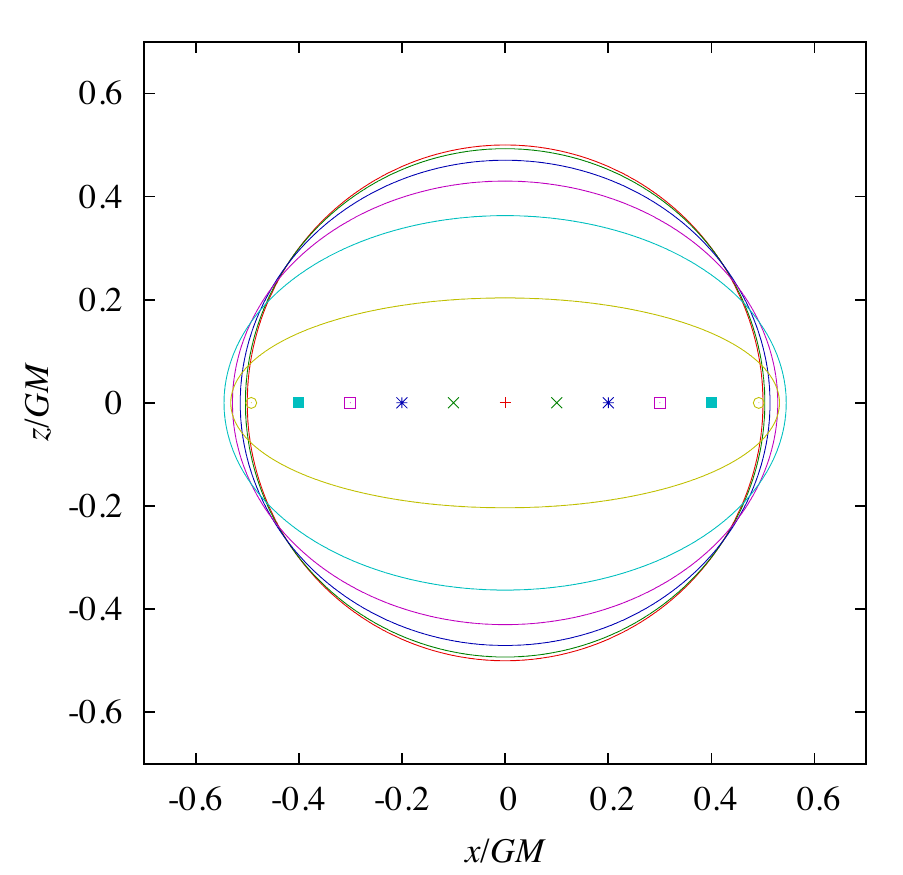}
\caption{Sections of marginally trapped spheres (apparent horizons
  and spherical minimal surfaces, at the same time) in ring systems 
  with the $(x,z)$-plane for $R/GM=0.1$, $0.2$, $0.3$,
  $0.4$, and $0.49203$.
  For $R/GM\ge 0.49204$, a marginally trapped sphere cannot be found.
  Compare with Fig.~\ref{figure_ring_DTTS_all}.
}
\label{figure_ring_spherical_AH_all}
\end{figure}
%

Figure~\ref{figure_ring_spherical_AH_all} depicts
a marginally trapped sphere (that is, an apparent horizon
and a minimal surface at the same time) for
values of $R/GM$ from $0.1$ to $0.4$ at $0.1$ intervals,
and for $0.49203$. For $R/GM= 0.49203$, the minimal surface 
approximately degenerates
with an inside maximal surface,
and they vanish for $R/GM\ge 0.49204$.\footnote{
  In Ref.~\cite{Chiba:1994}, the existence of an extremely oblate
  apparent horizon up to $R/GM\lesssim 0.70$ is reported. In 2001,
  H.Y. privately communicated with Takeshi Chiba, one of the
  authors of Ref.~\cite{Chiba:1994}, and we agreed that the extremely oblate
  apparent horizon could be a numerical artifact that appears
  when the resolution is not sufficient. We also agreed
  on the maximum value of $R/GM$ for the existence of an apparent horizon.
}

%
\begin{figure}[tb]
\centering
\includegraphics[width=0.7\textwidth,bb=0 0 360 162]{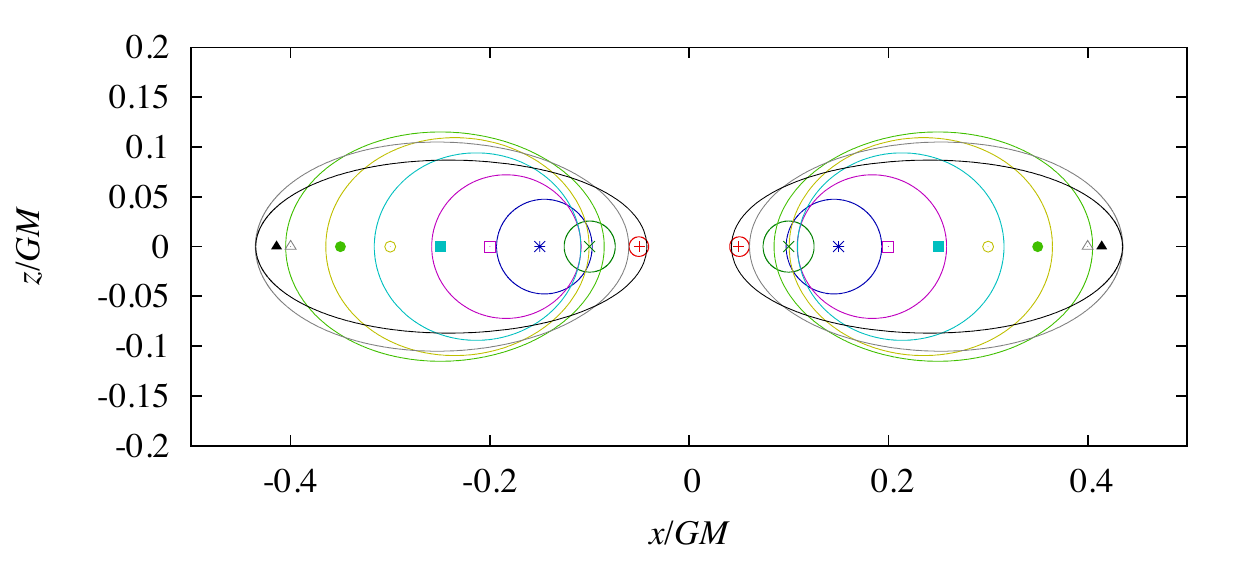}
\caption{Sections of marginally trapped tori (toroidal minimal surfaces,
  at the same time) in ring systems 
  with the $(x,z)$-plane for $R/GM=0.05$, $0.10$, $0.15$, $0.20$,
  $0.25$, $0.30$, $0.35$
  $0.40$, and $0.41413$.
  For $R/GM\ge 0.41414$, a marginally trapped torus cannot be found.
  Compare with Fig.~\ref{figure_ring_torus_DTTS_all}.
}
\label{figure_ring_torus_AH_all}
\end{figure}
%

%
\begin{figure}[tb]
\centering
\includegraphics[width=0.9\textwidth,bb=0 0 459 181]{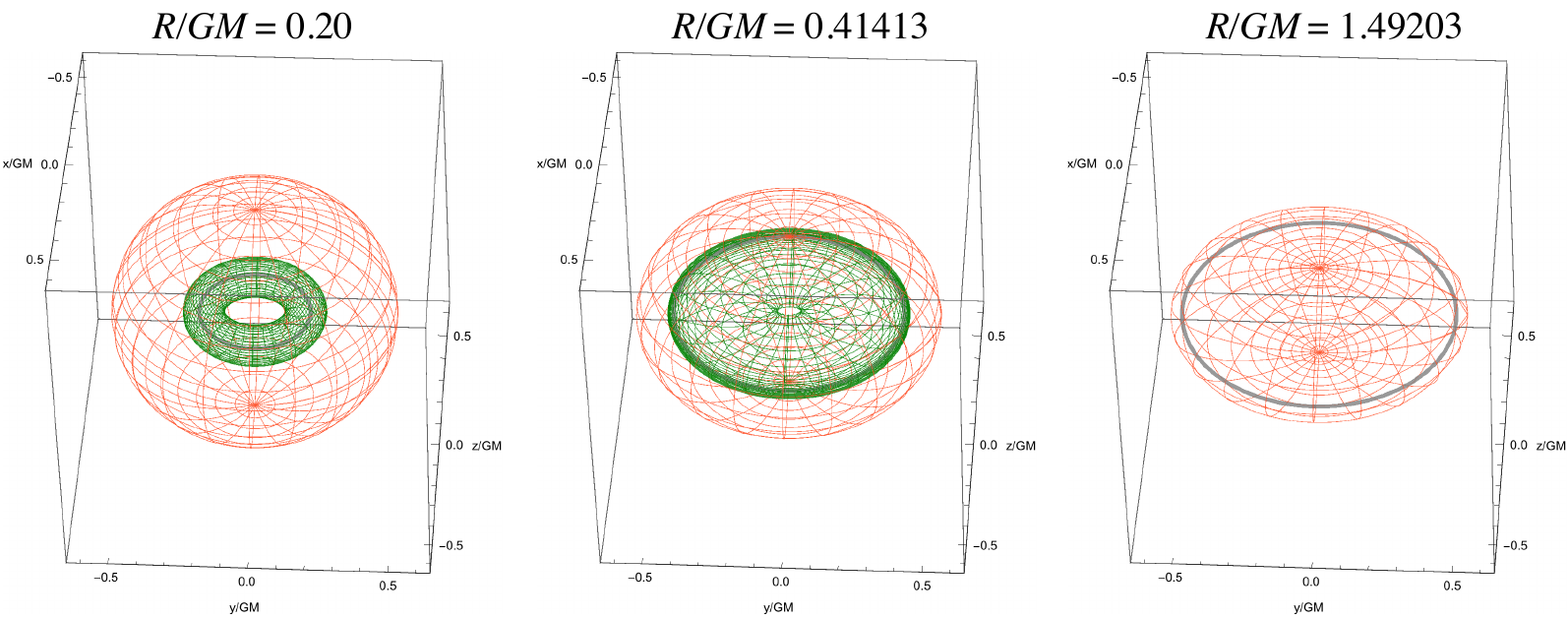}
\caption{Three-dimensional plots of
  a marginally trapped sphere and a marginally trapped torus
  for $R/GM=0.20$ (left panel),
  $0.41413$ (middle panel), and $0.49203$ (right panel).
  Compare with Fig.~\ref{figure_torus_spherical_TTS_radius}.
}
\label{figure_torus_spherical_AH_radius}
\end{figure}
%

Marginally trapped tori found in our numerical calculations
are plotted in Fig.~\ref{figure_ring_torus_AH_all}
for values of $R/GM$ from $0.05$ to $0.40$ at $0.05$ intervals,
and for $0.41413$. For $R/GM\ge 0.41414$, a marginally trapped torus
cannot be found. The marginally trapped torus is always located
inside the marginally trapped sphere. 
In Fig.~\ref{figure_torus_spherical_AH_radius},
the marginally trapped sphere and the marginally trapped torus
are plotted together for $R/GM=0.20$ (left panel),
$0.41413$ (middle panel), and $0.49203$ (right panel).
See also Appendix~\ref{Appendix-B} for an approximate analysis
for $R\ll GM$ that supports the existence of the marginally trapped torus.

%
\begin{figure}[tb]
\centering
\includegraphics[width=0.5\textwidth,bb=0 0 360 252]{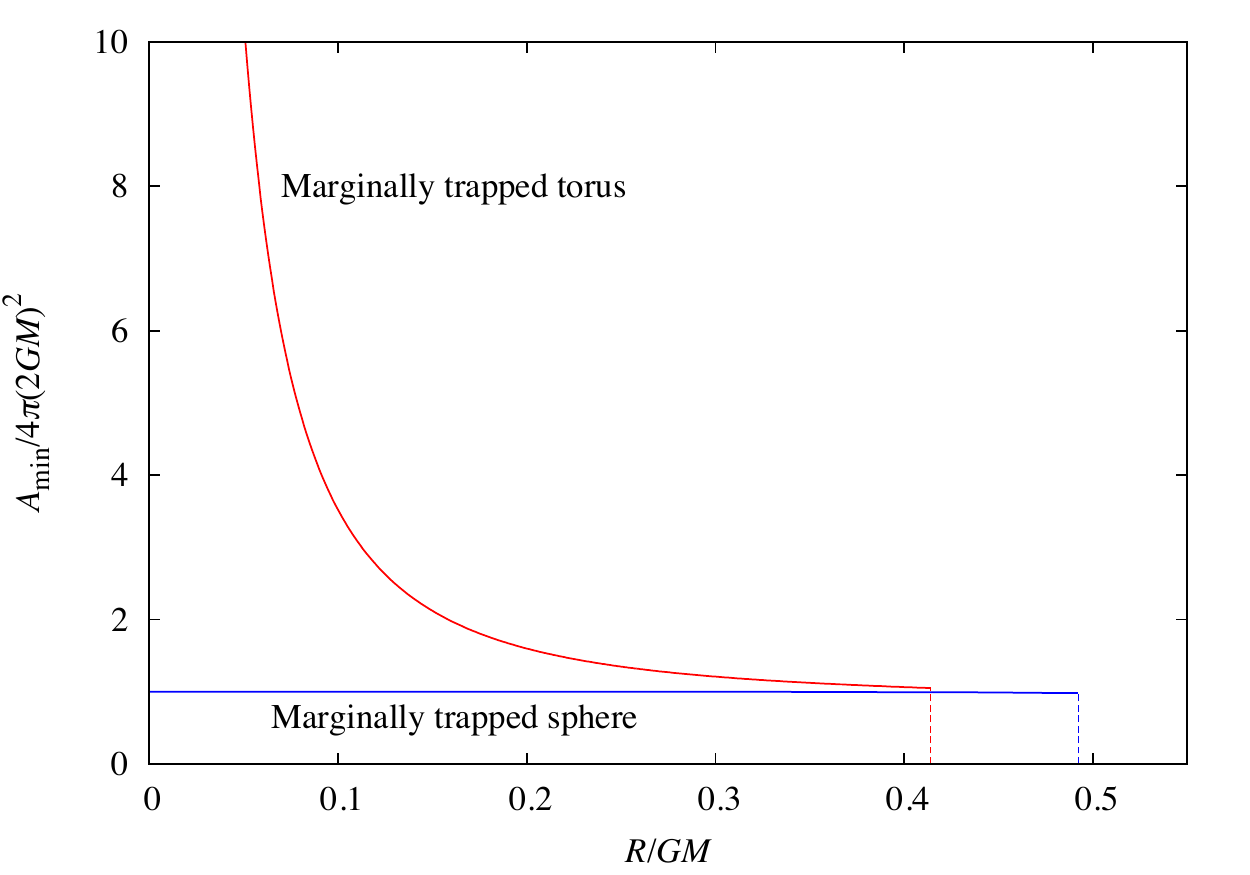}
\caption{The area $A_{\rm min}$ of a marginally trapped sphere and that of 
  a marginally trapped torus (a spherical minimal surface
  and a toroidal minimal surface, at the same time, respectively)
  as functions of $R/GM$.
  The area is normalized by $4\pi (2GM)^2$.
}
\label{figure_AH_area_ring}
\end{figure}
%

Figure~\ref{figure_AH_area_ring} shows the area of
the marginally trapped sphere and the marginally trapped torus
normalized by the horizon area in the spherically symmetric case,
$A_{\rm min}/4\pi(2GM)^2$ as functions of $R/GM$.
The area of the sphere monotonically decreases 
as $R/GM$ is increased,
and takes the value $A_{\rm min}/4\pi(2GM)^2\approx 0.9801$
at $R/GM=0.49203$. The area of the torus is also a monotonically
decreasing function of $R/GM$, and it is always greater than one.
As the value of $R/GM$ decreases to zero,
the area of the torus becomes unboundedly large.
Note that this result does not contradict the existing proofs 
of the Riemannian Penrose inequality
$A_{\rm min}\le 4\pi (2GM)^2$ \cite{Wald:1977,Huisken:2001,Bray:2001},
because those proofs apply only to the outermost minimal surface,
while in each of the present systems  
a toroidal minimal surface exists inside a spherical
maximal surface, which, in turn, exists inside a spherical
minimal surface.

%
%
\section{Summary and discussion}
\label{section6}

We have studied the formation of
marginally DTTSs in 
time-symmetric, conformally flat
initial data in order to examine whether
they can be understood analogously
to the hoop conjecture.
Three kinds of systems have been studied,
with two-black-hole initial data, spindle initial data,
and ring initial data.
In all systems, the condition $\mathcal{C}\lesssim 6\pi GM$
approximately describes the formation of the (outermost) 
marginally DTT sphere, and we have proposed
the stretched hoop conjecture for the formation of marginally DTT spheres
in Sect.~\ref{section4-3}.
Our results indicate that an arbitrarily long
DTT sphere is unlikely to form.

It has also been shown that in the ring system 
there is a parameter region of the ring radius $R$
where a marginally DTT torus forms.  
Such a marginally DTT torus is located inside
a marginally DTT sphere, consistent with the expectation
from the stretched hoop conjecture.
Since there is a parameter region where a marginally DTT torus
is not hidden inside an apparent horizon, marginally DTT tori
may be observable. 
In addition, there is a parameter region of $R$
where a marginally trapped torus forms inside a
marginally trapped sphere. 
This provides a further example of the similarity
between (marginally) DTTSs and (marginally) trapped surfaces.

The condition for the formation of marginally DTT tori
is left as a remaining problem. 
We point out that the same statement should also hold 
for marginally trapped tori. Although one may
expect that marginally DTT/trapped tori may form
if matter is concentrated in a ring-shaped region,
the existing works indicate that trapped tori
form even in the initial data
of a spherically symmetric star \cite{Karkowski:2017}.
Although the authors of Ref.~\cite{Karkowski:2017} discussed such a condition 
in terms of the binding energy of a star,
how to apply that condition to the present ring system
is unclear. 
Another remaining problem is that since a DTT torus has been shown to form,
DTTSs with the topologies of double torus, triple torus, and so on,
may form as well. In order to clarify this,
it is necessary to study non-axisymmetric initial data,
which will be more difficult compared to the study in this paper.

The present paper answers one of the remaining problems
listed in our previous paper \cite{Yoshino:2019}.
Since a (marginally) DTTS is a new 
concept, there still remain a lot of issues to be clarified,
i.e. the preparation of methods of solving for marginally DTTSs
on non-time-symmetric initial data,
the possible constraints from the presence of a DTTS on the global
properties of a spacetime, the connection to
a wandering set \cite{Siino:2019} that is the extension of a photon sphere
defined from global point of view, and exploring the connection
to observations. 
We hope to report on these in forthcoming papers.

\ack

H.Y. thanks Takeshi Chiba for helpful discussions in 2001.
H.Y. is supported by Grant-in-Aid for
Scientific Research (C) no. JP18K03654 from the Japan Society for
the Promotion of Science (JSPS).
K. I. is supported by JSPS Grant-in-Aid for Young Scientists (B)
no. JP17K14281.
T. S. is supported by Grant-in-Aid for Scientific Research (C) no. JP16K05344
from JSPS.
K.I. and T.S. are
also supported by Scientific Research (A) no. JP17H01091
and in part by JSPS Bilateral Joint Research Projects
(JSPS-NFR collaboration) ``String Axion Cosmology.''
The work of H.Y. is partly supported by
Osaka City University Advanced Mathematical Institute
(MEXT Joint Usage/Research Center on Mathematics and Theoretical Physics).

\appendix

%
%
\section{Equations for marginally DTT/trapped tori in ring initial data}
\label{Appendix-A}

In this appendix, we present the derivation
of the equations for marginally DTT/trapped tori
studied in Sects.~\ref{section3-2-2} and \ref{section5}.
Through the coordinate transformations
in Eqs.~\eqref{coordinate-transformation-tilderho-rho-xi}
and \eqref{coordinate-transformation-tildexi-rho-xi}, 
the metric in the neighborhood of $\sigma_0$ becomes
\begin{eqnarray}
  &ds^2\ =& ~~~\varPsi^4\left[1+(\rho+h)^2p_{,\rho}^2\right]d\rho^2
  \nonumber\\
  &&+\ \varPsi^4\left[h^{\prime 2}+(\rho+h)^2(1-p_{,\xi})^2\right]d\xi^2
  \nonumber\\
  &&+\ 2\varPsi^4\left[h^{\prime}-(\rho+h)^2(1-p_{,\xi})p_{,\rho}\right]d\rho d\xi
  \nonumber\\
  &&+\ \varPsi^4\left[R+(\rho+h)\cos(\xi-p)\right]^2d\phi^2,
\end{eqnarray}
and we require the coordinates $(\rho, \xi, \phi)$ to be orthogonal,
that is, 
\begin{equation}
h^{\prime}=(\rho+h)^2(1-p_{,\xi})p_{,\rho}.
\end{equation}
In particular, this relation means that 
\begin{equation}
  p_{,\rho} = \frac{h^\prime}{h^2}, \qquad
  p_{,\rho\xi} = \frac{h^{\prime\prime}}{h^2}-\frac{2h^{\prime 2}}{h^3}
\end{equation}
on $\sigma_0$. With respect to the orthonormal basis given by
Eq.~\eqref{orthonormal-basis-torus}, 
the diagonalized orthonormal components of the extrinsic curvature
defined in Eq.~\eqref{kab-diagonalized} are calculated from 
\begin{align}
k_{\xi\xi} = \frac{1}{2\varphi}\partial_\rho h_{\xi\xi}, \qquad & k_1=\frac{k_{\xi\xi}}{h_{\xi\xi}},\\
k_{\phi\phi} = \frac{1}{2\varphi}\partial_\rho h_{\phi\phi}, \qquad & k_2=\frac{k_{\phi\phi}}{h_{\phi\phi}}.
\end{align}
The result is
\begin{subequations}
\begin{eqnarray}
  k_1 &=& -\frac{h}{\varPsi^2(h^2+h^{\prime 2})^{3/2}}\left(h^{\prime\prime}+C\right),
  \label{k1-torus}
  \\
  k_2 &=& \frac{D}{\varPsi^2h\sqrt{h^2+h^{\prime 2}}},
  \label{k2-torus}
\end{eqnarray}
\end{subequations}
with 
\begin{subequations}
\begin{eqnarray}
  C&=& -h-2\frac{h^{\prime 2}}{h}
  -2\left(\frac{\varPsi_{,\tilde{\rho}}}{\varPsi}-\frac{h^\prime}{h^2}\frac{\varPsi_{,\tilde{\xi}}}{\varPsi}\right)
  (h^2+h^{\prime 2}),
  \label{torus-C}
  \\
  D&=& \frac{h\cos\xi+h^\prime\sin\xi}{R+h\cos\xi}h
  +2\left(\frac{\varPsi_{,\tilde{\rho}}}{\varPsi}-\frac{h^\prime}{h^2}\frac{\varPsi_{,\tilde{\xi}}}{\varPsi}\right)
  h^2,
  \label{torus-D}
\end{eqnarray}
\end{subequations}
where the relation
\begin{equation}
  \varPsi_{,\rho} = \varPsi_{,\tilde{\rho}}-\frac{h^\prime}{h^2}\varPsi_{,\tilde{\xi}}
\end{equation}
is used. The two-dimensional Ricci scalar ${}^{(2)}R$ on $\sigma_0$
that appears in Eq.~\eqref{Equation-for-a-marginally-DTTS-time-symmetric}
is calculated as
\begin{equation}
  \frac12{}^{(2)}R = \frac{1}{\varPsi^4(h^2+h^{\prime 2})}(Ah^{\prime\prime}+B),
  \label{Ricci-scalar-torus}
\end{equation}
with
\begin{subequations}
\begin{equation}
  A=\frac{2h^\prime}{h^2+h^{\prime 2}}
  \frac{\varPsi_{,\tilde{\rho}}h^\prime +\varPsi_{,\tilde{\xi}}}{\varPsi}
  -2\frac{\varPsi_{,\tilde{\rho}}}{\varPsi}
  -\frac{h\cos\xi+h^\prime\sin\xi}{(R+h\cos\xi)(h^2+h^{\prime 2})}h,
  \label{torus-A}
\end{equation}
\begin{multline}
  B =
  2\left(\frac{\varPsi_{,\tilde{\rho}}h^\prime +\varPsi_{,\tilde{\xi}}}{\varPsi}\right)^2
  -\frac{2}{\varPsi}\left(\varPsi_{,\tilde{\rho}\tilde{\rho}}h^{\prime 2}
  +2\varPsi_{,\tilde{\rho}\tilde{\xi}}h^{\prime} 
  +\varPsi_{,\tilde{\xi}\tilde{\xi}}\right)
  +\frac{h\cos\xi+h^\prime\sin\xi}{(R+h\cos\xi)(h^2+h^{\prime 2})}
  (h^2+2h^{\prime 2})
  \\
  +\frac{2}{(R+h\cos\xi)(h^2+h^{\prime 2})}
  \frac{\varPsi_{,\tilde{\rho}}h^\prime +\varPsi_{,\tilde{\xi}}}{\varPsi}
  \left[h^3\sin\xi-h^{\prime 3}\cos\xi+hh^{\prime}(R+h^\prime \sin\xi)\right],
  \label{torus-B}
\end{multline}
\end{subequations}
where the relations
\begin{subequations}
\begin{eqnarray}
  \varPsi_{,\xi} &=& \varPsi_{,\tilde{\rho}}h^\prime+\varPsi_{,\tilde{\xi}},
  \\
  \varPsi_{,\xi\xi}&=& \varPsi_{,\tilde{\rho}}h^{\prime\prime}
  +\varPsi_{,\tilde{\rho}\tilde{\rho}}h^{\prime 2}
  +2\varPsi_{,\tilde{\rho}\tilde{\xi}}h^{\prime}
  +\varPsi_{,\tilde{\xi}\tilde{\xi}}
\end{eqnarray}
\end{subequations}
are used. We now present the equations for marginally
DTT/trapped tori in terms of $h$, $A$, $B$, $C$, and $D$.

\subsection{Marginally DTT tori with $k_1\le k_2$}

The equations for marginally DTT tori must be studied
for the cases $k_1\le k_2$ and $k_1\ge k_2$, separately. 
In the case $k_1\le k_2$, Eq.~\eqref{Equation-for-a-marginally-DTTS-time-symmetric}
becomes
\begin{equation}
    h^{\prime\prime} = \frac{-2CD+(D^2/h^2-B)(h^2+h^{\prime 2})}{2D+A(h^2+h^{\prime 2})}. 
  \label{equation-prolate-case}
\end{equation}

\subsection{Marginally DTT tori with $k_1\ge k_2$}

In the case $k_1\ge k_2$, Eq.~\eqref{Equation-for-a-marginally-DTTS-time-symmetric}
becomes
\begin{multline}
  h^{\prime\prime} =
  -C + \frac{h^2+h^{\prime 2}}{h^2}\left[D+\frac12(h^2+h^{\prime 2})A\right]
  \\
  \mp \frac{h^2+h^{\prime 2}}{h^2}\sqrt{\left[D+\frac12(h^2+h^{\prime 2})A\right]^2+h^2(B-AC)}.
  \label{equation-oblate-case}
\end{multline}
This equation includes a double sign. In the case of marginally DTT
spheres, it is possible to choose an appropriate sign
by requiring the equation
to be consistent with the presence of a photon sphere
in the spherically symmetric case \cite{Yoshino:2019}. Since an appropriate
sign cannot be chosen from such physical considerations
in the toroidal case, we study the cases for both signs.
As a result, a marginally trapped torus can be obtained
for the minus sign of Eq.~\eqref{equation-oblate-case}.
For the plus sign, we could not obtain a solution satisfying
the boundary conditions.

\subsection{Marginally trapped tori}

Since a marginally trapped
surface coincides with a minimal surface in time-symmetric initial data,
the equation for marginally trapped tori becomes $k=k_1+k_2=0$. 
This is equivalent to
\begin{equation}
h^{\prime\prime} = -C+\frac{h^2+h^{\prime 2}}{h^2}D.
\end{equation}

%
%
\section{Approximate analysis of marginally DTT/trapped tori}
\label{Appendix-B}

In this appendix, we proceed with an approximate analysis
of the marginally DTT/trapped tori in the ring initial data 
in order to provide evidence for their existence and
support the numerical results in Sect.~\ref{section5}. 
We focus on the case $GM\gg R$, for which our numerical results indicate 
that the radii of marginally DTT/trapped tori become much 
smaller compared to $R$. For this reason,
we consider the expanded form of $h(\tilde{\xi})$
of Eq.~\eqref{parametrization-torus} as
\begin{equation}
 h(\tilde{\xi}) = \rho_0+\sum_{n\ge 1}\rho_n\cos (n\tilde{\xi}), 
\end{equation}
and assume 
\begin{equation}
R \gg \rho_0\gg \rho_n \quad (n=1,2, \ldots).
\end{equation}
In the approximation, 
we ignore the terms with $O(R/M)$, $O(\rho_0/R)$, $O(\rho_n/\rho_0)$
compared to the leading terms.
We substitute $a+b \approx 4R^2$ into 
Eq.~\eqref{Psi-ring-final-form} for the conformal factor,
and approximate the complete elliptic integral of the first kind
$K(\kappa)$ 
with the expansion formula around $\kappa = 1$ up to order one, 
\begin{equation}
K(\kappa) = -\frac12\log(1-\kappa^2)+2\log 2+O((1-\kappa^2)\log(1-\kappa^2)),
\end{equation}
which is derived
using {\it Mathematica}.\footnote{
  We have also rigorously proved this relation through calculations by hand.}
Here, $1-\kappa^2$ is approximated as
$1-\kappa^2 \approx {\tilde{\rho}^2}/{4R^2}$, and  
then, we have
\begin{equation}
  \varPsi \ \approx\ -\frac{GM}{2\pi R}\log\left(\frac{\tilde{\rho}}{8R}\right).
\end{equation}
From the formulae presented in Appendix~\ref{Appendix-A}, 
the two-dimensional Ricci scalar ${}^{(2)}R$
is ignored because ${}^{(2)}R\ll O(\varPsi^{-4}\rho_0^{-2})$
holds, where $\varPsi^{-4}\rho_0^{-2}$ is the order of $k_1^2$ and $k_1k_2$.
Equations \eqref{torus-C} and \eqref{torus-D} for
$C$ and $D$, which are related to $k_1$ and $k_2$ through
Eqs.~\eqref{k1-torus} and \eqref{k2-torus}, are approximated as
\begin{eqnarray}
  C&\approx& -h - 2\frac{\varPsi_{,\tilde{\rho}}}{\varPsi}h^2
  \ \approx\ -\rho_0\left[1+\frac{2}{\log({\rho_0}/{8R})}\right],
  \\
  D&\approx& 2\frac{\varPsi_{,\tilde{\rho}}}{\varPsi}h ^2
  \ \approx\ \frac{2\rho_0}{\log({\rho_0}/{8R})}.
\end{eqnarray}
We now solve for marginally DTT tori and marginally trapped
tori.

\subsection{Marginally DTT tori}

Because $k_1\ge k_2$ holds, 
the equation for marginally DTT tori becomes $k_1+2k_2\approx 0$,
and this is equivalent to $C\approx 2D$. Solving this equation, we have
\begin{equation}
\rho_0 \ \approx\ 8\exp(-6) R \ \approx\ 0.01983\times R,
\end{equation}
and hence, a solution consistent with the assumption $\rho_0\ll R$ is obtained.
Let us compare this result with the 
numerically obtained data. For the parameter $R/M = 1.0\times 10^{-3}$,
the numerical result shows that the radius 
takes values in the range 
$1.974\times 10^{-5} \lesssim h(\tilde{\xi})/M\lesssim 2.014\times 10^{-5}$
due to the angular dependence of $h(\tilde{\xi})$.
This is consistent with the radius from the
approximation, $\rho_0/M\approx 1.983\times 10^{-5}$. 

\subsection{Marginally trapped tori}

The equation for marginally trapped tori is $k_1+k_2=0$,
and this is equivalent to $C\approx D$. Solving this equation, we have
\begin{equation}
\rho_0 = 8\exp(-4) R \ \approx\ 0.1465\times R.
\end{equation}
Although the value of
$\rho_0/R$ is relatively large and the approximation may not be very good,
let us proceed further.
For the parameter $R/M = 1.0\times 10^{-3}$,
the numerical result shows that the radius 
takes values in the range 
$1.359\times 10^{-4} \lesssim h(\tilde{\xi})/M\lesssim 1.570\times 10^{-4}$,
which is consistent with the radius from the
approximation, $\rho_0/M\approx 1.465\times 10^{-4}$.

\end{document}